\documentclass[twocolumn]{aastex61}

\newcommand\aastex{AAS\TeX}

\newcommand{\Agama}{\textsc{Agama }}

\received{}
\revised{}
\accepted{}

\shorttitle{\aastex\ Peculiar radial velocities of RSG candidates in RSGC4}
\shortauthors{Chun et al.}

\begin{document}

\title{Is the RSGC4 (Alicante 8) cluster a real star cluster?: peculiar radial velocities of red supergiant stars}

\correspondingauthor{Sang-Hyun Chun}
\email{shyunc@kasi.re.kr}
\author[0000-0002-6154-7558]{Sang-Hyun Chun}
\affiliation{Korea Astronomy and Space Science Institute, 776 Daedeokdae-ro, Yuseong-gu, Daejeon 34055, Republic of Korea} 
\author[0000-0002-5629-8876]{GyuChul Myeong}
\affiliation{Institute of Astronomy, University of Cambridge, Madingley Road, Cambridge, CB3 0HA, UK}
\author[0000-0003-0894-7824]{Jae-Joon Lee}
\affiliation{Korea Astronomy and Space Science Institute, 776 Daedeokdae-ro, Yuseong-gu, Daejeon 34055, Republic of Korea}
\author[0000-0002-0418-5335]{Heeyoung Oh}
\affiliation{Korea Astronomy and Space Science Institute, 776 Daedeokdae-ro, Yuseong-gu, Daejeon 34055, Republic of Korea}



\begin{abstract}
Young massive star clusters, like the six red supergiant clusters in the Scutum complex, provide valuable insights into star-formation and galaxy structures.
We investigated the high-resolution near-infrared spectra of 60 RSG candidates in these clusters using the Immersion Grating Infrared Spectrograph. 
Among the candidates in RSGC4, we found significant scattering in radial velocity ($-64$ km/s to $115$ km/s),
unlike other clusters with velocities of $\sim$100 km/s.
Most candidates in RSGC4 have $Q_{GK_s}$ values larger than 1.7, suggesting that they could be early AGB stars.
Four candidates in RSGC4 exhibit infrared excess and distinct absorption features absent in other candidates.
Two of these stars exhibit absorption lines resembling those of D-type symbiotic stars, showing radial velocity changes in multi-epoch observations.
Analysis of relative proper motions revealed no runaway/walkaway stars in RSGC4.
The dynamic properties of RSGC4 and RSGC1 differ from the disk-like motions of other clusters: RSGC4 has low normalized
horizontal action $J_\mathrm{hor}=J_\mathrm{\phi}/J_\mathrm{tot}$ and vertical action $J_\mathrm{ver}=(J_\mathrm{z}-J_\mathrm{R})/J_\mathrm{tot}$ values and high eccentricities, 
while RSGC1 has vertical motions with high $J_\mathrm{ver}$ values and inclinations.
We propose that RSGC4 may not be a genuine star cluster but rather a composite of RSGs and AGBs distributed along the line of sight at similar distances, possibly originating from various environments.
Our results suggest a complex and hierarchical secular evolution of star clusters in the Scutum complex, emphasizing the importance of considering factors beyond density crowding when identifying star clusters in the bulge regions.
\end{abstract}




\section{Introduction}\label{sec:intro}
Young massive star clusters (YMCs) containing high-mass stars are ideal laboratories 
for investigating the various astrophysical aspects of galaxies.
YMCs are intrinsically bright and are considered the basic building blocks of galaxies.
Most stars form within star clusters~\citep{Lada2003}, and YMCs, in particular, contain substantial populations of massive stars~\citep{Zinn2007}.
YMCs also lose members to galaxy field through dissolution processes, such as dynamical evolution~\citep[e.g.][]{Poveda1967,Gies1986,Fujii2011,Oh2015} or tidal stripping~\citep[e.g.][]{Tutukov1978,Kroupa2001,Baumgardt2008}.
Therefore, YMCs are powerful tracers of star formation, evolutionary history, and parent galaxy structures. 

YMCs also play a distinct role as probes for studying stellar evolution and dynamics in star clusters.
YMCs contain large populations of massive, coeval stars~\citep{Zinn2007} with similar chemical compositions, from which
we can trace recent massive star formations and study their stellar evolution.

Many massive stars are found in multiple systems and undergo interactions with companion stars throughout their lifetimes~\citep{Sana2012,Kobul2014,Dunstall2015,Moe2017}.
Entire stellar evolutionary paths can be changed by binary interactions.
Processes such as binary supernova ejection~\citep{Blaauw1961,Renzo2019} or dynamical encounters~\citep{Poveda1967,Banerjee2012,Perets2012,Oh2016} in multiple systems can transfer stars from their parent clusters to galaxy fields. Thus, 
the binary systems in clusters significantly influence the dynamical evolution of the clusters.
A deeper understanding of the properties of binary systems in YMCs is crucial for comprehending their binary effects on massive stellar evolution 
and the dynamic evolution of clusters.

Despite their significant contributions to star formation, galaxy evolution, and cluster dynamics,
our understanding of the properties of YMCs remains incomplete.
This is because YMCs with an $M_{\mathrm{cluster}} >10^4$ $M_{\sun}$ are very scarce in the Milky Way, and massive stars in the clusters are also rare due 
to their fast evolution.
Fortunately, recent near-infrared all-sky surveys, such as 2MASS and WISE, have proven invaluable for discovering many obscured star clusters 
in the Milky Way~\citep[][]{Ivanov2002,Camargo2016,Ryu2018}.

The discovery of six red supergiant clusters (RSGCs) at the intersection of the Scutum-Crux arm and the Galactic bar tip (hereafter, referred to as the Scutum 
complex region) in the Milky Way ($l\sim24$--29$\arcdeg$) is extremely interesting.
These clusters include RSGC1~\citep{Figer2006}, RSGC2~\citep{Davies2007}, 
RSGC3~\citep{Clark2009}, RSGC4~\citep[Alicante 8;][]{Negu2010}, RSGC5~\citep[Alicante 7;][]{Negu2011}, and Alicante 10~\citep{Gon2012}.
These discoveries not only allow a detailed investigation of massive star evolution but also raise many intriguing questions about star formation in this region.

RSGCs all have large populations of RSGs, with 10--25 RSGs in each cluster and an age range of 12--20 Myr. 
The presence of such large numbers of RSGs indicates that the initial mass range needed for clusters to harbor such stars is in the range of 2--4$\times10^4 M_{\sun}$.
The close proximity of clusters, with projected separation ranges of 31--400 pc at almost identical distances of 5--6 kpc, and the presence of a diffuse field of RSGs~\citep{Garzon1997,Lopez1999} in the Scutum complex region
suggest a recent starburst resulting from interactions between the Galactic bar and the spiral arm~\citep{Garzon1997,Davies2007}.
Physical associations between clusters, which are similar to those observed in the extended cluster complexes of external galaxies~\citep[e.g., M51,][]{Bastian2005}, are also expected to exist in the Scutum complex region.

Interestingly, subsolar metallicities have been reported for these clusters~\citep{Davies2009,Origlia2016,Origlia2019,Chun2020}, 
indicating the presence of low-metallicity gas fueling into the inner disks~\citep{Magrini2009,Bono2013}. 
This is inconsistent with the high metallicity properties of this area,
and is not considered in current Galactic evolution models. 
Therefore, it is important to characterize the cluster parameters through follow-up observations.

The recent starburst, potential cluster associations, and subsolar metallicities observed in these clusters 
motivated us to obtain high-resolution spectra for studying the dynamical properties of RSGs in six RSGCs in the Scutum complex.
In this study, we refer to RSGs as RSG candidates, as their confirmation requires high-resolution spectroscopy.
We note that high-resolution spectroscopic studies have not been conducted solely for RSGC4.
Recently,~\citet{Asad2023} conducted a study on RSGC4 using spectroscopic data (R$\sim$5600) from VLT/X-shooter.
In Section 2, we introduce the sample selection, observation, and data reduction. In Section 3, we describe the peculiar radial velocities of the RSG candidates in RSGC4
and those in the other RSGCs. In Section 4, we explore various potential explanations for the peculiar radial velocities of RSG candidates 
in RSGC4. Finally, we present our conclusions and a summary in Section 5.

\section{Target selection, Observation, and data reduction}
\begin{table*}
{\tiny
\begin{center}
\caption{The observation log, radial velocities of observed RSG candidates and proper motions from $Gaia$ DR3\label{kinematics}}
\begin{tabular}{ccccccccc}
\hline
ID & RA & DEC & Exp. time (s) & S/N & $V_{Helio}$ (km s$^{-1}$) & $V_{LSR}$ (km s$^{-1}$) & $\mu_\alpha$cos$\delta$ (mas yr$^{-1}$)   &  $\mu_\delta$ (mas yr$^{-1}$) \\
\hline
RSGC4 (12) &   &  &   &  & $\overline{V}_{Helio}$=7.3$\pm65.4$ & $\overline{V}_{LSR}$=22.6$\pm65.6$ & $\overline{\mu_\alpha cos\delta}$=-1.731$\pm1.757$ & $\overline{\mu_\delta}$=-4.942$\pm1.927$ \\ 
RSGC4-1                 & 18:34:58.40 & -07:14:24.8 & 640  & 260  & -28.7$\pm$0.7  & -12.9$\pm$0.7     & -0.662$\pm$0.269 & -1.986$\pm$0.229     \\
RSGC4-1\tablenotemark{a}                 &                     &                     & 160  & 200  & -25.4$\pm$0.9     & -10.2$\pm$0.9     &                               &                                   \\
RSGC4-1\tablenotemark{b}                 &                     &                     &         & 174  & -33$\pm$0.5        &                             &                               &     \\
RSGC4-2\tablenotemark{a}                 & 18:34:55.12 & -07:15:10.8  & 200  & 210  & 73.7$\pm$0.3      & 88.9$\pm$0.3      & -1.854$\pm$0.351 & -2.695$\pm$0.300            \\
RSGC4-2\tablenotemark{b}                 &                     &                     &         &  92   & 71$\pm$0.6         &                             &                               &    \\
RSGC4-3                                             & 18:34:50.00 & -07:14:26.2  & 240  & 380  & 22.0$\pm$0.3      & 37.8$\pm$0.3      & -0.964$\pm$0.277 & -5.726$\pm$0.232             \\
RSGC4-3\tablenotemark{a}                 &                     &                     & 200  & 410  & 20.6$\pm$0.3      & 35.9$\pm$0.3      &                               &               \\
RSGC4-3\tablenotemark{b}                 &                     &                     &         & 146  & 37$\pm$0.5         &                             &                               &                \\
RSGC4-4\tablenotemark{a}                 & 18:34:51.02 & -07:14:00.5 & 200  & 380   & 90.8$\pm$0.4      & 106.1$\pm$0.4    & -1.032$\pm$0.815 & -5.108$\pm$0.706              \\
RSGC4-4\tablenotemark{b}                 &                     &                     &        &  43    & 88$\pm$0.7         &                             &                                &           \\
RSGC4-5                 & 18:34:51.33 & -07:13:16.3 & 200  & 300  & -10.8$\pm$0.4     & 5.0$\pm$0.4        & -3.276$\pm$0.297 & -7.537$\pm$0.253              \\
RSGC4-5\tablenotemark{a}                               &  &  & 200  & 270  & -12.3$\pm$0.4     & 3.1$\pm$0.4        &  &               \\
RSGC4-5\tablenotemark{b}                 &                     &                     &         & 102   & -9$\pm$0.5         &                             &                                &          \\
RSGC4-6                 & 18:34:41.55 & -07:11:38.8 & 960  & 190  & 114.9$\pm$0.4    & 130.7$\pm$0.4    & -2.127$\pm$0.462 & -6.021$\pm$0.398             \\
RSGC4-6\tablenotemark{a}                                &  &  & 200  & 310  & 113.3$\pm$0.2     & 128.6$\pm$0.2    &  &              \\
RSGC4-6\tablenotemark{b}                 &                     &                     &         & 112   & 111$\pm$0.6        &                            &                                 &          \\
RSGC4-7                 & 18:34:43.56 & -07:13:29.7 & 320  & 320  & -2.0$\pm$0.2      & 13.3$\pm$0.2       & -3.831$\pm$0.288 & -6.550$\pm$0.246             \\
RSGC4-7\tablenotemark{b}                 &                     &                     &         & 110   & -4$\pm$0.6          &                            &                                  &            \\
RSGC4-8                 & 18:34:44.51 & -07:14:15.3 & 640  & 260  & -0.1$\pm$0.7      & 15.7$\pm$0.7       & -0.676$\pm$0.263 & -2.859$\pm$0.230             \\
RSGC4-8\tablenotemark{a}                               &  &  & 200  & 150  & -5.0$\pm$0.3      & 10.2$\pm$0.3       &  &              \\
RSGC4-8\tablenotemark{a}                                &  &  & 52   & 173  & -1.5$\pm$0.3       & 13.7$\pm$0.3       &  &              \\
RSGC4-9                 & 18:34:45.81 & -07:18:36.2 & 960  & 240  & -64.4$\pm$0.7   & -48.6$\pm$0.7      & 1.116$\pm$0.273 & -6.562$\pm$0.217              \\
RSGC4-9\tablenotemark{a}                                &  &  & 320  & 360  & -81.1$\pm$0.2    & -65.9$\pm$0.2      &  &               \\
RSGC4-10                & 18:34:26.81 & -07:15:27.9 & 192  & 210  & 1.0$\pm$0.6      & 16.2$\pm$0.6      & 0.324$\pm$0.889 & -1.270$\pm$0.721              \\
RSGC4-11                & 18:35:00.32 & -07:07:37.4 & 200  & 280  & 44.8$\pm$0.4    & 60.6$\pm$0.4      & -1.636$\pm$0.249 & -3.933$\pm$0.212              \\
RSGC4-11\tablenotemark{a}                                 &  &  & 400   & 310  & 45.8$\pm$0.6    & 61.1$\pm$0.6     &  &               \\
RSGC4-12\tablenotemark{a}                & 18:35:16.88 & -07:13:26.9 & 240  & 330  & -13.5$\pm$1.2   & 1.7$\pm$1.2       & -5.325$\pm$0.309 & -6.766$\pm$0.260               \\
\hline
RSGC1 (14)&   &  &   &  & $\overline{V}_{Helio}$=105.9$\pm2.6$ & $\overline{V}_{LSR}$=122.0$\pm3.2$ & $\overline{\mu_\alpha cos\delta}$=-3.961$\pm0.478$ & $\overline{\mu_\delta}$=-4.832$\pm0.800$ \\
RSGC1-1                 & 18:37:56.29 & -06:52:32.2 & 300  & 430  & 107.2$\pm$1.0   & 122.5$\pm$1.0   & -4.569$\pm$0.717 & -5.074$\pm$0.603              \\
RSGC1-2                 & 18:37:55.28 & -06:52:48.4 & 200  & 490  & 100.5$\pm$0.7   & 115.7$\pm$0.7    & -4.065$\pm$0.792 & -6.729$\pm$0.658               \\
RSGC1-3                 & 18:37:59.72 & -06:53:49.4 & 280  & 590  & 110.8$\pm$0.7   & 126.1$\pm$0.7    & -2.945$\pm$0.728 & -4.865$\pm$0.596               \\
 &  &  &  &  &  & &  &  \\
\hline
RSGC2 (11) &   &  &   &  & $\overline{V}_{Helio}$=90.3$\pm3.0$ & $\overline{V}_{LSR}$=106.9$\pm3.1$ & $\overline{\mu_\alpha cos\delta}$=-2.181$\pm0.267$ & $\overline{\mu_\delta}$=-4.627$\pm0.219$ \\
RSGC2-13                  & 18:39:17.7  & -06:04:02.5 & 120  & 390  & 93.0$\pm$0.5      & 110.0$\pm$0.5    & -2.101$\pm$0.227 & -4.739$\pm$0.196               \\
RSGC2-16                & 18:39:24.0  & -06:03:07.3 & 120  & 390  & 90.5$\pm$0.5      & 107.6$\pm$0.5    & -2.015$\pm$0.124 & -4.269$\pm$0.101               \\
RSGC2-17                & 18:39:15.1  & -06:05:19.1 & 120  & 350  & 89.7$\pm$0.5      & 106.7$\pm$0.5    & -2.052$\pm$0.206 & -4.695$\pm$0.175               \\
 & &  &  &  &  &  &  &  \\
\hline
RSGC3 (9)&   &  &   &  & $\overline{V}_{Helio}$=91.0$\pm12.5$ & $\overline{V}_{LSR}$=107.5$\pm12.5$ & $\overline{\mu_\alpha cos\delta}$=-2.348$\pm0.641$ & $\overline{\mu_\delta}$=-5.016$\pm0.812$ \\
RSGC3-1                 & 18:45:23.60 & -03:24:13.9 & 150  & 310  & 92.3$\pm$0.4      & 108.7$\pm$0.4    & -2.012$\pm$0.214 & -5.903$\pm$0.166               \\
RSGC3-6                 & 18:45:19.39 & -03:24:48.3 & 240  & 320  & 106.3$\pm$0.4    & 122.7$\pm$0.4    & -2.590$\pm$0.278 & -5.934$\pm$0.225               \\
RSGC3-8                 & 18:45:20.06 & -03:22:47.1 & 320  & 320  & 89.8$\pm$0.5      & 106.2$\pm$0.5    & -2.561$\pm$0.230 & -5.385$\pm$0.188               \\
 & &  &  &  &  &  &  &  \\

\hline
RSGC5 (6) &   &  &   &  & $\overline{V}_{Helio}$=77.0$\pm23.0$ & $\overline{V}_{LSR}$=93.4$\pm23.0$ & $\overline{\mu_\alpha cos\delta}$=-1.824$\pm0.340$ & $\overline{\mu_\delta}$=-4.458$\pm0.902$  \\
RSGC5-1                 & 18:44:20.53 & -03:28:44.6 & 320  & 410  & 79.6$\pm$0.5     & 96.1$\pm$0.5     & -1.686$\pm$0.175 & -4.166$\pm$0.145               \\
RSGC5-4                 & 18:44:27.96 & -03:29:42.5 & 100  & 370  & 86.0$\pm$0.5     & 103.4$\pm$0.5    & -2.360$\pm$0.219 & -4.749$\pm$0.185               \\
RSGC5-5                 & 18:44:29.45 & -03:30:02.4 & 80   & 310  & 84.1$\pm$0.5      & 100.5$\pm$0.5    & -1.493$\pm$0.213 & -4.191$\pm$0.189               \\
& &  &  &  &  &  &  &  \\

\hline
Alicante10 (8) &   &  &   &  & $\overline{V}_{Helio}$=70.8$\pm11.1$ & $\overline{V}_{LSR}$=87.2$\pm11.1$ & $\overline{\mu_\alpha cos\delta}$=-2.269$\pm0.860$ & $\overline{\mu_\delta}$=-5.081$\pm1.052$ \\
Alican10-1                & 18:45:07.94  & -03:41:24.0 & 360 & 380  & 66.6$\pm$0.4    & 83.0$\pm$0.4     & -0.747$\pm$0.354 & -4.769$\pm$0.313              \\
Alican10-2               & 18:45:01.85  & -03:42:06.5 & 300  & 260  & 76.9$\pm$0.6    & 93.3$\pm$0.6     & -2.998$\pm$0.469 & -5.628$\pm$0.346             \\
Alican10-3               & 18:45:17.60  & -03:43:05.1 & 180  & 310  & 80.3$\pm$0.5    & 96.7$\pm$0.5     & -3.106$\pm$0.290 & -6.142$\pm$0.234              \\
& &  &  &  &  &  &  &  \\
\hline
\end{tabular}
\end{center}
}
\tablecomments{}{Full Table~\ref{kinematics} is published in the machine-readable format.
A portion is shown here for guidance regarding its form and content. Redundant information is eliminated.
The first header of each cluster shows average radial velocity and proper motions. The numbers in parentheses represent the counts
of observed RSG candidates in each cluster.}
\tablenotetext{a}{Radial velocities are estimated from the spectra obtained in 2021.}
\tablenotetext{b}{Radial velocities from~\citet{Asad2023}.}
\end{table*}

We selected RSG candidates in six young massive star clusters from several published catalogs: 
16 stars in RSGC~\citep{Davies2008}, 26 stars in RSGC2~\citep{Davies2007},
15 stars in RSGC3~\citep{Clark2009}, 13 stars in RSGC4~\citep{Negu2010}, 9 stars in RSGC5~\citep{Negu2011}, and 8 stars in Alicante 10~\citep{Gon2012}.
In the case of RSGC2, the stars whose chemical abundances were not investigated by~\citet{Davies2009} were selected as the observation targets.
The distribution of RSG candidates in the clusters in $(J-K_s, K_s)$ color-magnitude diagram was examined, and we preselected as many target stars as possible.

High-resolution near-infrared spectra of the RSG candidates in the six star clusters were obtained using the Immersion Grating Infrared Spectrometer (IGRINS) installed
as a visitor instrument on the Gemini South telescope. 
The IGRINS simultaneously provides high-resolution (R$\sim$45,000) spectra of the $H$ and $K$ bands, covering a wavelength range from
14,800 to 24,800 $\mathrm{\AA}$, for which we used a $0.34''\times5''$ slit.
Detailed and extensive explanations of IGRINS have been provided by~\citet{Yuk2010}, ~\citet{Park2014}, and ~\citet{Mace2018}.

Observations were conducted by two poor weather programs in service mode in June and July 2018 (Program ID: GS-2018A-Q-418 for RSGC2, RSGC3, RSGC4, RSGC5, and Alicante 10) and in July and August 2021 (Program ID: GS-2021A-Q-417 for RSGC1 and RSGC4) because our targets are so bright.
Most of RSG candidates were secured using the ABBA nod sequence, while some stars were observed using the ON-OFF sequence, depending on weather conditions.
Individual exposure times were varied between 20 and 180 s to obtain high signal-to-noise ratios (median S/N$ > $100 over the entire spectra of the
$H$ and $K$ bands in the final reduced spectra).
In addition to the scientific targets, we obtained A0V telluric standard stars at air masses close to those of the scientific targets 
during the same night to remove the telluric absorption lines. The observation logs are listed in Table~\ref{kinematics}.

The data reduction process was performed using the IGRINS reduction pipeline package~\citep{Lee2017}. 
This pipeline facilitates standard reduction processes, including bad pixel correction, flat fielding, geometric transformation, background removal, 
wavelength calibration, and order extraction.
The resulting spectra were divided by the A0V spectra to remove the telluric features and then multiplied by a model spectrum of Vega.
The final S/N ratios per pixel of the reduced spectra, representing the median S/N across the entire spectrum, are provided in Table~\ref{kinematics}.

\section{Peculiar radial velocities of red supergiant stars in RSGC4} 
\begin{figure*}
\includegraphics[width=0.7\textwidth]{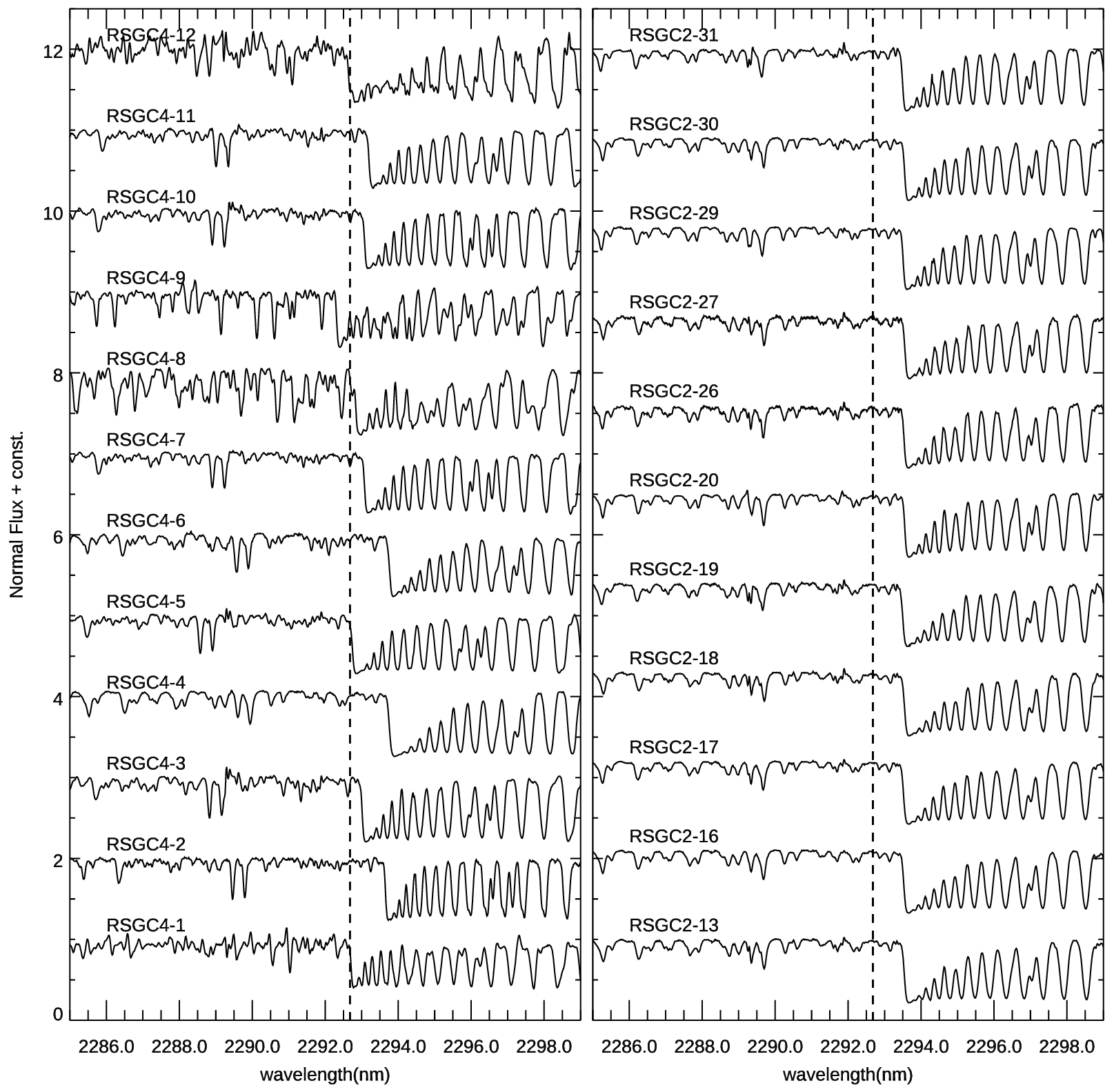}
\centering
\caption{Left: IGRINS $K$-band spectra with only heliocentric correction for all observed stars in RSGC4 in the region around the CO bandhead. 
The ID numbers of the stars correspond with those of the stars in RSGC4 by~\citet{Negu2010}.
Right: $K$-band spectra resulting with only heliocentric correction for observed stars in RSGC2 in this study.
The ID numbers for RSGC2 follow those of~\citet{Davies2007}.
The dotted vertical line indicates the rest wavelength position of the blue edge of the CO bandhead. 
}\label{Kspec}
\end{figure*}

Figure~\ref{Kspec} shows the IGRINS spectra for the region around the CO bandhead in the $K$-band of the observed stars in 
RSGC4 (left panel) and RSGC2 (right panel). Only heliocentric correction was applied to the spectra.
The ID numbers of the stars in both panels correspond to those of RSGC4 of~\citet{Negu2010} and RSGC2 of~\citet{Davies2007}.
All spectra exhibit strong absorption features and deep CO bandhead absorption, which are characteristic of late-type
stars. 
The amount of Doppler shift for stars is shown qualitatively with respect to the rest wavelength of the blue edge of the CO bandhead (dotted line) in Figure~\ref{Kspec}.
In the left panel of Figure~\ref{Kspec}, we find blue-shifted or red-shifted blue edges of the CO bandhead in the RSGC4 spectra. The extent of the shift is also inconsistent. 
In contrast, the wavelengths at the CO bandhead of the RSGC2 spectra are nearly identical.
This discrepancy suggests that the RSG candidates in RSGC4 do not have consistent radial velocities. 

To accurately determine the radial velocities of the stars, the radial velocities for all observed RSG candidates in the six clusters were 
estimated by cross-correlation of the spectra with synthetic RSG spectra.
In the high-resolution IGRINS spectra, several absorption lines of the RSG candidates enable us to conduct precise radial velocity estimations. 
We selected several spectral segments containing prominent absorption lines, such as Ti, Mg and CO bands in the $H$ and $K$ bands to compare the synthetic RSG spectra.

Synthetic spectra for the $H$ and $K$ bands were calculated using local thermodynamic equilibrium (LTE) line analysis and synthetic
spectrum code MOOG~\citep{Sneden1973}. We used a MARCS model atmosphere~\citep{Gustaf2008} with typical stellar parameters for an RSG 
(i.e., $T_{eff}=3800$ K, log $g$=0.0, and metallicity of $\mathrm{[Fe/H]}=-0.21$). We note that several RSGs in RSGCs are known to have subsolar metallicity~\citep{Davies2008,Origlia2013}.
Atomic and molecular line lists for computing the synthetic spectra were adopted from~\citet{Afsar2016}, who combined atomic and CO molecular lines from
~\citet{Kurucz2011} and CN and OH molecular lines from~\citet{Sneden2014} and~\citet{Brooke2016}.  
We estimated the radial velocities of the spectral segments and averaged them to obtain the final radial velocity.
Standard deviations of the average radial velocities were used to represent velocity errors.
The heliocentric radial velocities and velocities in the local standard of rest (LSR) reference system were calculated from
the final radial velocities using the {\it rvcorrect} task in IRAF~\citep{Tody1986,Tody1993}.

\begin{figure*}
\includegraphics[width=\textwidth]{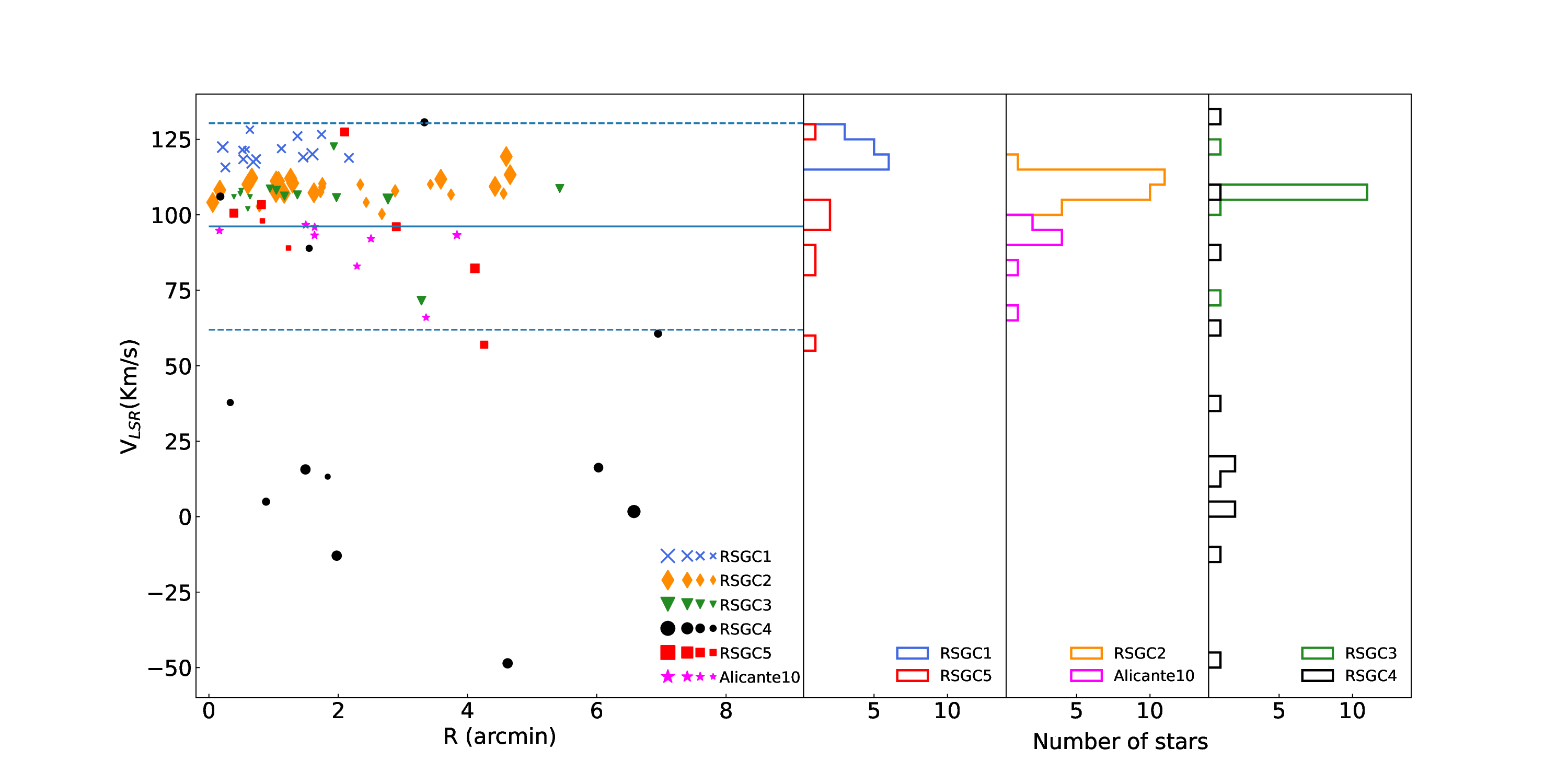}
\caption{Radial velocity distributions as a function of radial distance from the cluster center in the sky for six RSGCs. 
Symbol size represents the measured radial velocity error of an individual star, with larger symbols indicating larger errors.
Four symbols corresponding to different velocity error categories ($\sigma_{V_{LSR}}=4.0, 1.0, 0.6$, and $0.3$ km s$^{-1}$) are adopted.
The solid line indicates the mean radial velocity ($V_{LSR}=96$ km s$^{-1}$) for all RSG candidates in the clusters. 
The dashed lines represent the $1\sigma$ (34 km s$^{-1}$) range from the mean velocity.
Histograms of radial velocity for each cluster are also presented in the right-hand three panels.
}\label{radial}
\end{figure*}
The derived radial velocities of the RSG candidates in each cluster are indicated in Table~\ref{kinematics}. 
As predicted in Figure~\ref{Kspec}, the radial velocity values of the RSG candidates in the RSGC4
exhibit significant variations, spanning a wide range from $-64$ to $115$ km s$^{-1}$. The mean velocity is 18.98 km s$^{-1}$, with a standard deviation of 50.26 km s$^{-1}$.
We note that this substantial velocity difference among the RSG candidates in RSGC4 was not found by~\citet{Negu2010}. Although they did not estimate the radial velocities of the RSGs, 
the Doppler shifts in their spectra appear almost identical~\citep[see Figure 4 in][]{Negu2010}.
We attribute this discrepancy to the lower resolution (R$\sim$2500) of their spectra.
We note that we observed the same RSG candidates identified by~\citet{Negu2010}. 
There is no chance of observing different stars because they are the most luminous objects in this field. 
~\citet{Chun2023} reported inconsistent radial velocities for RSG candidates in RSGC4.
~\citet{Asad2023} also demonstrated a large dispersion in the radial velocities of RSG candidates in RSGC4, suggesting that RSGC4 may not
be a genuine cluster.

The radial velocities of the RSG candidates in the RSGC4 were compared with those of the other RSGCs.
Previous medium-resolution spectroscopic studies~\citep{Davies2007,Davies2008,Negu2011} suggested that 
all RSGCs in the Scutum complex were located within $5\sim6$ kpc from us, potentially sharing a common radial velocity of 95--123 km s$^{-1}$.
For a proper comparison, we collected the radial velocities of the RSG candidates that were not included in our observations from the previous 
catalogs. These values were added to our radial velocity measurements.
Specifically, we incorporated radial velocity data from~\citet{Davies2008} for RSGC1,~\citet{Davies2007} for RSGC2,~\citet{Origlia2016} for
RSGC3, and~\citet{Origlia2019} for RSGC5 and Alicante10.

Figure~\ref{radial} shows the radial velocity distributions in the LSR for the RSG candidates in the six RSGCs as a function of the radial distance from each cluster center in the sky. 
The median values for the sky positions of the RSG candidates in the clusters were assigned to the cluster center. 
Symbol size in the plot corresponds to the velocity error, with larger symbols indicating larger errors.
We adopted four symbols to represent different velocity error categories: $\sigma_{V_{LSR}}=4.0, 1.0, 0.6$, and $0.3$ km s$^{-1}$.
Additionally, histograms of the radial velocity distributions for the candidates in each cluster are presented in the right-hand three panels.

As previously reported, all RSG candidates in the five clusters (excepting RSGC4)
have common radial velocities within the range of 75--125 km s$^{-1}$, suggesting that they are probably members of the Scutum complex.
However, the RSG candidates in RSGC4 show very different radial velocity distributions compared to the other clusters.
Most of the RSG candidates in RSGC4 have low radial velocities, with a large spread. RSGC4-1 and RSGC4-9 even exhibit
negative $V_{LSR}$ velocities.

\section{Discussion of the peculiar radial velocities of RSG candidates in RSGC4}
The unusual radial velocities observed for the RSG candidates within RSGC4 raise many intriguing questions regarding their membership and the underlying cause of these anomalous velocities.
The most straightforward explanation for the peculiar radial velocities of the RSG candidates in RSGC4 is that some of the identified stars may belong to different populations or represent a random overdensity of field stars rather than true members of the star cluster.
Another consideration is the potential influence of binaries or multiplicity in star clusters on radial velocity. The third interpretation involves the presence of unbounded stars 
in a cluster.
A final possibility is that environmental perturbations might have been responsible for the peculiar radial velocities.
We discuss these four interpretations to better understand the peculiar radial velocities observed for the RSG candidates in RSGC4, including the RSG candidates in the other five clusters.

\begin{table*}
{\scriptsize
\begin{center}
\caption{The photometry and color information of RSG candidates in six RSGCs\label{photo}}
\begin{tabular}{ccccccccccccccc}
\hline
  & \multicolumn{3}{c}{\rm 2MASS}  & $Gaia$ & \multicolumn{4}{c}{GLIMPSE}  & \multicolumn{2}{c}{DENIS} & \multicolumn{2}{c}{DENIS-2MASS}  & $Q1$ & $Q2$  \\
\cline{2-15} 
  ID & $J$ & $H$ & $K_s$ & $G$ & [3.6] & [4.5] & [5.8] &[8.0] & $J$ & $K_s$ &  $\Delta$ $J$ & $\Delta$ $K_s$  &  &  \\
\hline
RSGC4-1      & 9.921  & 7.796 & 6.671 & 17.685 & 6.646  & 6.017  & 5.536  & 5.321  & 10.316 & 7.028  & 0.396  & 0.358  & 0.086  & -0.379 \\
RSGC4-2      & 10.86  & 8.452 & 7.324 & 18.828 & 6.691  & 6.731  & 6.367  & 6.308  & 10.765 & 7.374  & 0.095  & 0.054  & 0.376  & 0.818  \\
RSGC4-3      & 9.551  & 7.186 & 5.995 & 17.658 & 5.500  & 5.926  & 4.844  & 4.679  & 9.672  & 6.157  & 0.122  & 0.157  & 0.218  & -0.003 \\
RSGC4-4      & 10.731 & 7.675 & 6.141 & 20.023 & 5.218  & 5.268  & 4.653  & 4.495  & 10.875 & 6.320   & 0.145  & 0.180   & 0.278  & 0.165  \\
RSGC4-5      & 10.037 & 7.527 & 6.264 & 18.433 & \nodata & \nodata & \nodata & \nodata & 10.209 & 6.467  & 0.169  & 0.207  & 0.224  & \nodata \\
RSGC4-6      & 10.255 & 7.731 & 6.429 & 19.065 & 6.632  & 6.090  & 5.300  & 5.188  & 10.473 & 6.718  & 0.213  & 0.288  & 0.190   & 0.489  \\
RSGC4-7      & 10.456 & 8.224 & 7.055 & 18.494 & 6.837  & 6.426  & 5.942  & 5.877  & 10.664 & 7.317  & 0.204  & 0.257  & 0.152  & 0.218  \\
RSGC4-8      & 10.416 & 8.030  & 6.623 & 18.019 & \nodata & 6.513  & 4.754  & 4.625  & 10.527 & 6.894  & 0.107  & 0.274  & -0.041 & -1.460 \\
RSGC4-9      & 9.829  & 7.629 & 6.486 & 17.567 & 5.384  & 5.327  & 4.993  & 4.831  & \nodata & \nodata & \nodata & \nodata & 0.148  & -1.123 \\
RSGC4-10     & 10.984 & 8.319 & 6.997 & 19.775 & 6.712  & 6.181  & 5.780   & 5.605  & 11.218 & 7.356  & 0.238  & 0.356  & 0.284  & 0.227  \\
RSGC4-11    & 9.610  & 7.469 & 6.342 & 17.452 & 5.688  & 6.135  & 5.349  & 5.118  & \nodata & 6.642  & \nodata & 0.302  & 0.106  & -0.017 \\
RSGC4-12     & 9.872  & 7.728 & 6.454 & 18.120  & 5.267  & 4.998  & 4.655  & 4.448  & 9.833  & 6.493  & 0.037  & 0.043  & -0.164 & -1.965 \\
\hline
RSGC1-1      & 9.748  & 6.587 & 4.962 & 19.516 & \nodata & \nodata & \nodata & \nodata & 9.857  & 4.252  & 0.109  & 0.710  & 0.236  & \nodata \\
RSGC1-2      & 9.904  & 6.695 & 5.029 & 19.549 & \nodata & 3.585  & 2.736  & \nodata & 9.933  & 4.330  & 0.029  & 0.699  & 0.210  & \nodata \\
RSGC1-3      & 9.954  & 6.921 & 5.333 & 19.219 & 4.301  & \nodata & 3.704  & \nodata & 9.952  & 4.628  & 0.002  & 0.705  & 0.175  & \nodata \\
  &    &   &   &  &    &    &    &  &    &    &    &    &   &  \\
\hline
RSGC2-13     & 8.421  & 6.387 & 5.439 & 15.320  & 5.415  & 5.008  & 4.587  & 4.382  & 8.506  & 4.688  & 0.085  & 0.751  & 0.328  & 0.139  \\
RSGC2-16     & 8.235  & 6.444 & 5.597 & 14.447 & 5.561  & \nodata & 4.870  & 4.762  & 8.218  & 4.578  & 0.017  & 1.019  & 0.266  & 0.392  \\
RSGC2-17     & 8.709  & 6.613 & 5.619 & 15.854 & 7.177  & 6.718  & \nodata & \nodata & 8.717  & 4.741  & 0.008  & 0.878  & 0.307  & \nodata \\
&    &   &   &  &    &    &    &  &    &    &    &    &   &  \\\hline
RSGC3-1      & 8.551   & 6.539  & 5.578  & 15.379 & \nodata & 6.019  & 4.726  & 4.499  & \nodata & \nodata & \nodata & \nodata & 0.282  & 0.062  \\
RSGC3-6      & 9.062   & 6.968  & 6.036  & 16.130  & \nodata & 6.246  & 5.243  & 5.199  & \nodata & \nodata & \nodata & \nodata & 0.416  & 0.758  \\
RSGC3-8      & 9.525  & 7.288  & 6.303   & 16.746 & 6.789  & 6.249  & 5.462  & 5.430  & \nodata & \nodata & \nodata & \nodata & 0.458  & 0.890   \\
&    &   &   &  &    &    &    &  &    &    &    &    &   &  \\
\hline
RSGC5-1      & 9.280   & 7.235  & 6.213  & \nodata & 5.657  & 6.085  & 5.381  & 5.445  & 9.156  & 6.100    & 0.124  & 0.110   & 0.186  & 1.012  \\
RSGC5-4      & 8.063   & 5.910  & 4.851  & 15.182 & \nodata & 4.308  & 3.843  & \nodata & 7.981  & 4.142  & 0.079  & 0.708  & 0.242  & \nodata \\
RSGC5-5      & 8.446   & 6.386  & 5.409  & 15.386 & 4.794  & 5.148  & 4.567  & 4.447  & 8.335  & 5.010   & 0.115  & 0.400    & 0.296  & 0.450   \\
&    &   &   &  &    &    &    &  &    &    &    &    &   &  \\
\hline
Alicante10-1 & 10.500   & 8.080  & 6.910  & 18.712 & 6.637  & \nodata & 5.658  & 5.268  & \nodata & \nodata & \nodata & \nodata & 0.314  & -0.827 \\
Alicante10-2 & 10.457  & 7.943  & 6.707   & 18.672 & 6.651  & 6.342  & 5.644  & 5.576  & 10.334 & 6.687  & 0.116  & 0.013  & 0.278  & 0.726  \\
Alicante10-3 & 9.970   & 7.405   & 6.062  & 18.405 & 5.016  & \nodata & 4.761  & 4.521  & \nodata & \nodata & \nodata & \nodata & 0.158  & -0.230  \\
&    &   &   &  &    &    &    &  &    &    &    &    &   &  \\
\hline
\end{tabular}
\end{center}
}
\tablecomments{}{Full Table~\ref{photo} is published in the machine-readable format.}
\end{table*}

\subsection{Contamination with AGB stars}
Asymptotic giant branch stars (AGBs) in the field can lead to significant scattering in the radial velocity distribution.
The separation of AGBs and RSGs based solely on photometry has been a very difficult task.
AGBs have similar red colors to those of RSGs, and their luminosities also overlap with those of faint RSGs: $M_{\mathrm{bol}}$ from
$-5.5$ to $-9.0$ mag for RSGs~\citep{Davies2007,Clark2009}, and $M_{\mathrm{bol}}$ from $-5.5$ to $-7.0$ mag for AGBs~\citep{Alard2001,Messineo2014}.
In spectroscopic data, strong and broad water absorption features at the edges of the $H-$band and $K-$band spectra are used to
distinguish between AGBs and RSGs~\citep[e.g.,][]{Messineo2014}. AGB spectra typically show a highly curved continuum due to $\mathrm{H_{2}O}$ in their envelopes~\citep[e.g.,][]{Lancon2000}.
Unfortunately, our high-resolution IGRINS spectra of the RSG candidates were not appropriate for detecting continuum absorption by water, as
the continuum shape in high-resolution spectra is unreliable, and the broad water vapor feature spans multiple orders in the spectra.

Instead, we used the photometric variations of AGBs, especially Mira-AGBs, as a way to distinguish between AGBs and RSGs. 
AGBs generally have larger variable amplitudes than RSGs. The visual amplitude of a Mira-AGB exceeds 2.5 mag, extending up to $\sim1$ mag in $K_s$~\citep{Messi2004}.
In contrast, the variable amplitude of RSGs is a few tenths of the magnitude in the $K$ band~\citep{Kiss2006,Groe2009,Yang2011}.
~\citet{Messineo2014} conducted tests on the variability of the RSGs and AGBs using data from the Two Micron All-Sky Survey~\citep[2MASS;][]{Cutri2003} and 
the Deep Near-Infrared Survey~\citep[DENIS;][]{Epch1999}.
They identified variable Mira-AGBs in the Milky Way based on criteria such as $J(DENIS)>7.5$ mag, and $|J(DENIS)-J(2MASS)|>3\times0.15$ mag, or
$K_s(DENIS)>6.0$ mag, and $|K_s(DENIS)-K_s(2MASS)|>3\times0.15$ mag.
We followed these criteria to check the variability of RSG candidates in RSGC4 using the 2MASS and DENIS magnitudes.
The photometric magnitudes of 2MASS and DENIS are listed in Table~\ref{photo}.
We found that none of the RSG candidates in RSGC4 were consistent with these criteria, except for RSGC4-9, for which
the criteria could not be applied due to the lack of a DENIS counterpart.
This indicates that the RSGs candidates in RSGC4 do not show the level of variability observed in the Mira-AGBs.

\begin{figure*}
\includegraphics[width=\textwidth]{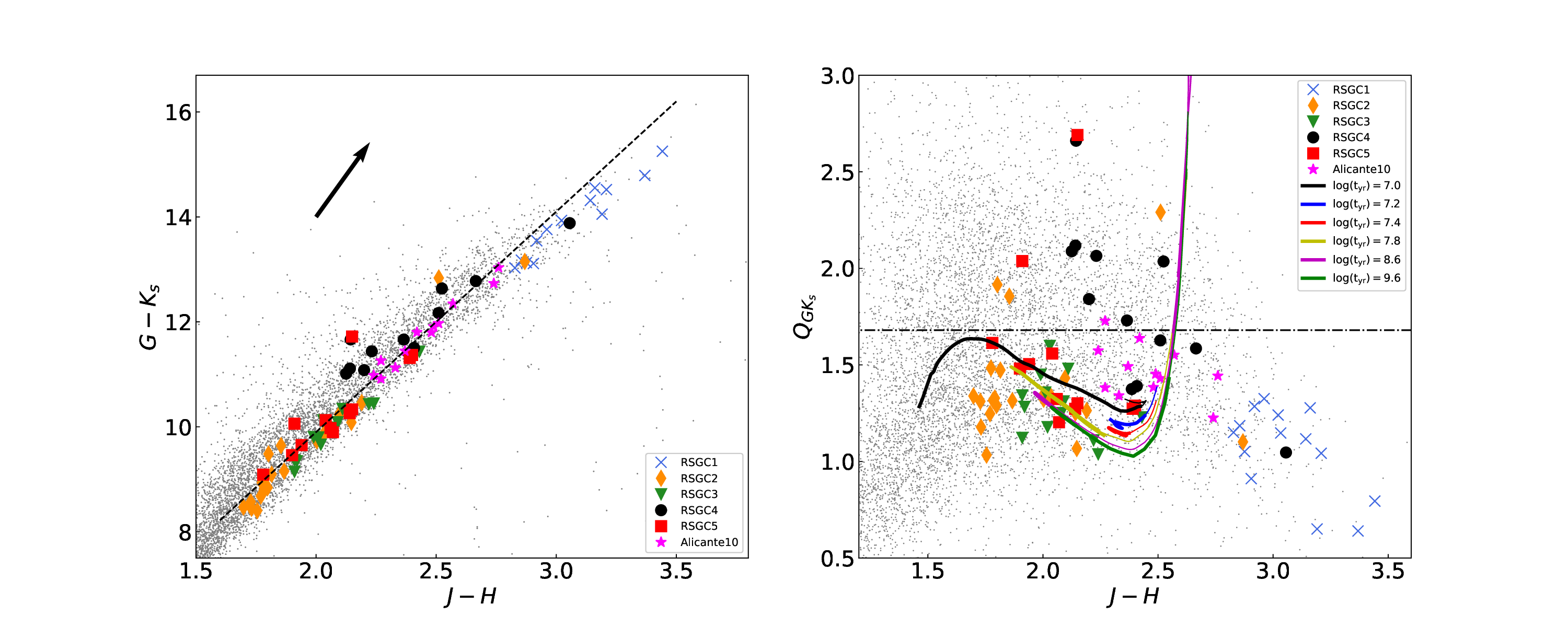}
\caption{Left: $(J-H, G-K_s)$ color-color diagram of stars with $K_s<8$ in the $6^\circ\times2^\circ$ Scutum complex area and RSG candidates in the six RSGCs. 
The arrow indicates an increasing reddening vector, and the dashed line shows the linear distribution of stars. Right: $(J-H, Q_{GK_s})$ diagram for the stars in the left panel. The MIST isochrones with [Fe/H]=$-0.25$ and log$\mathrm{(t_{yr})}=7.0, 7.2, 7.4, 7.8, 8.6$ and $9.6$ are indicated by solid lines of different colors. 
The core helium burning (CHeB) phases in isochrones are described by bold lines, while early AGB or thermal pulsing AGB phases are shown by thin lines. 
The dash-dotted line indicates the limit $Q_{GK_s}$ value of 1.7, which stars in the CHeB cannot attain.
}\label{selection}
\end{figure*}

We also studied the variability of RSG candidates in RSGC4 using the unTimely Catalog~\citep{Meisner2023}, which is based on data
from the Wide-field Infrared Survey Explorer~\citep[WISE;][]{Wright2010} and the NEOWISE mission extension~\citep{Mainzer2011,Mainzer2014}. 
 We found amplitudes of variation ranging from 0.2 to 1.0 mag in the W1 and W2 bands of WISE/NEOWISE single exposure photometry.
In the case of unTimely photometry, which is much deeper and
cleaner than WISE/NEOWISE single exposure photometry,
the amplitudes of variation were less than 0.5 mag in both the W1 and W2 bands.
It is worth noting that the expected amplitudes of variation of AGB stars typically exceeds 2 mag in both the W1 and W2 bands~\citep[e.g.,][]{Suh2021,Groen2022}.
Unfortunately, however, several RSG candidates in RSGC4 were saturated during certain epochs, potentially resulting in underestimated amplitudes of variation.
Therefore, although we could not infer definite variability similar to AGB stars for RSG candidates in RSGC4, the detected amplitudes of variation in WISE/NEOWISE data represent a lower limit of variation.  
Additional time-domain survey data in near-infrared, where RSG candidates are not saturated, are necessary to confirm the contamination of the Mira-AGBs in RSGC4.
We note that the VISTA Variables in the Via Lactea (VVV) survey~\citep{Minniti2010} did not cover the area of RSGC4, preventing us from investigating their variability.

We further investigated the contamination with field AGBs in RSGC4 using Gaia Data Release 3~\citep[$Gaia$ DR3;][]{Gaiadr3} and 2MASS data. 
The magnitudes in $G$ band of $Gaia$ data are shown in Table~\ref{photo}.
The left panel of Figure~\ref{selection} shows the
$(J-H, G-K_s)$ color-color diagram (CCD) for stars with $K_s<8$ in the $6^\circ\times2^\circ$ area where all six RSGCs are located. 
All RSG candidates in the six RSGCs are indicated with the same symbols as in Figure~\ref{radial}. 
We find that all the stars closely follow the reddening line (indicated by an arrow) in the $(J-H, G-K_s)$ CCD.
The RSG candidates in the six RSGCs are also located based on their distances and increasing extinctions with distance. 
The extinction toward RSGC4 appears to be higher than those toward RSGC2, RSGC3, and RSGC5, but lower than that toward RSGC1.
Therefore the distance to RSGC4 is somewhere between RSGC2 and RSGC1. 
This is consistent with previous distance and extinction results for RSGCs: $A_{K_s}=2.6$ and $d=6.6$ kpc for RSGC1~\citep{Davies2008}, 
$A_{K_s}=1.44$ and $d=5.8$ kpc for RSGC2~\citep{Davies2007}, $A_{K_s}=1.5$ and $d=6.0$ kpc for RSGC3~\citep{Clark2009},
$A_{K_s}=1.9$ and $d=6.6$ kpc for RSGC4~\citep{Negu2010}, and $A_{K_s}=1.5$ and $d=6.1$ kpc for RSGC5~\citep{Negu2011}.
Alicante 10 and RSGC4 appear to be at the same distance and have the same extinction value, as shown in Figure~\ref{selection}. 
~\citet{Gon2012} reported 6.0 kpc and $E(J-K_s)=2.6$, corresponding to $A_{K_s}=1.93$, for Alicante 10.
We note that the distances to and extinctions toward the clusters can differ by up to 1 kpc and $\Delta A_{K_s}=0.2$, respectively, 
depending on the methods used for the distance and extinction measurements.

Based on the stellar distributions and the linear fitting line (dashed line) in the $(J-H, G-K_s)$ CCD, 
we defined a new parameter, $Q_{GK_s}=(G-K_s)-4.24\pm0.01\times(J-H)$.
This parameter is independent of reddening, similar to the 
$Q1$\footnote{$Q1=(J-H)-1.8\times(H-K_s)$} parameter of~\citet{Messi2012}. 
The right panel of Figure~\ref{selection} shows the $Q_{GK_s}$ parameter
versus the $(J-H)$ diagram of stars and RSG candidates in the left panel. 
The MIST isochrones~\citep{Choi2016} with [Fe/H]=$-0.25$ and several ages of $\mathrm{log(t_{yr})}$=7.0, 7.2, 7.4, 7.8, 8.6, and 9.6 are also indicated 
to describe the stellar evolutionary phase. 
For clarity, only the isochrones from core helium burning (CHeB, bold lines) to 
early AGB or thermally pulsing AGB (EAGB and TPAGB, thin lines) phases are included in the plot.

The isochrones indicate that the stars in the CHeB phase do not surpass a specific $Q_{GK_s}$ value (i.e., 1.7 in this study). 
The majority of RSG candidates in the five RSGCs, except RSGC4, have $Q_{GK_s}$ values below 1.7, which
is consistent with the isochrone trend.
We note that only six stars in three clusters (three in RSGC2, two in RSGC5, and one in Alicante 10) have $Q_{GK_s}> 1.7$.
Three stars in RSGC2 are spectroscopically confirmed members~\citep[ID-2, ID-8, and ID-52;][]{Davies2007}. 
The two stars in RSGC5 are the main RSG and faint star~\citep[ID-A7 and ID-A12;][]{Negu2011}, respectively.
The star in Alicante 10 has been spectroscopically confirmed as an RSG~\citep[ID-Near07,][]{Gon2012}.
Unfortunately, data for these six stars were not obtained during the IGRINS observation; thus,
we did not investigate their spectroscopic properties in this study. 

Conversely, more than half RSG candidates in RSGC4 have $Q_{GK_s}$ values larger than $1.7$.
The isochrones indicate that the stars in the EAGB with $\mathrm{log(t_{yr})}=7.8$ or the TPAGB phase with $\mathrm{log(t_{yr})}=8.6$ and $9.6$ 
can have large $Q_{GK_s}$ values. If we adopted TPAGBs with $\mathrm{log(t_{yr})}=8.6$ and $9.6$ for the stars with $Q_{GK_s}>1.7$,
the distance to and foreground extinction toward RSGC4 should be equal to or less than those of RSGC2. 
However, as clearly shown in the $(J-H, G-K_s)$ CCD, the distance and extinction of RSGC4 are greater than those of RSGC2.
We find that the EAGBs with $\mathrm{log(t_{yr})}=7.8$ can partially explain the red colors and bright magnitudes
of stars in RSGC4 by adjusting the distance and extinction values ($d$=6.28 kpc and $A_{K_s}=1.69$), with only slight changes.
We note that the isochrones from the updated PARSEC code~\citep{Bressan2012} also yield the same results for the $Q_{GK_s}$ values of the 
RSG candidates in RSGC4. This indicates that the stars with $Q_{GK_s}>1.7$ in RSGC4 may not be
CHeB stars (except RSG with high mass-loss and thick circumstellar dust), but instead be foreground EAGB stars with initial masses of 
at least 6.48 $\mathrm{M_\odot}$ at almost the same distance as RSGC4.

\begin{figure}
\includegraphics[width=\columnwidth]{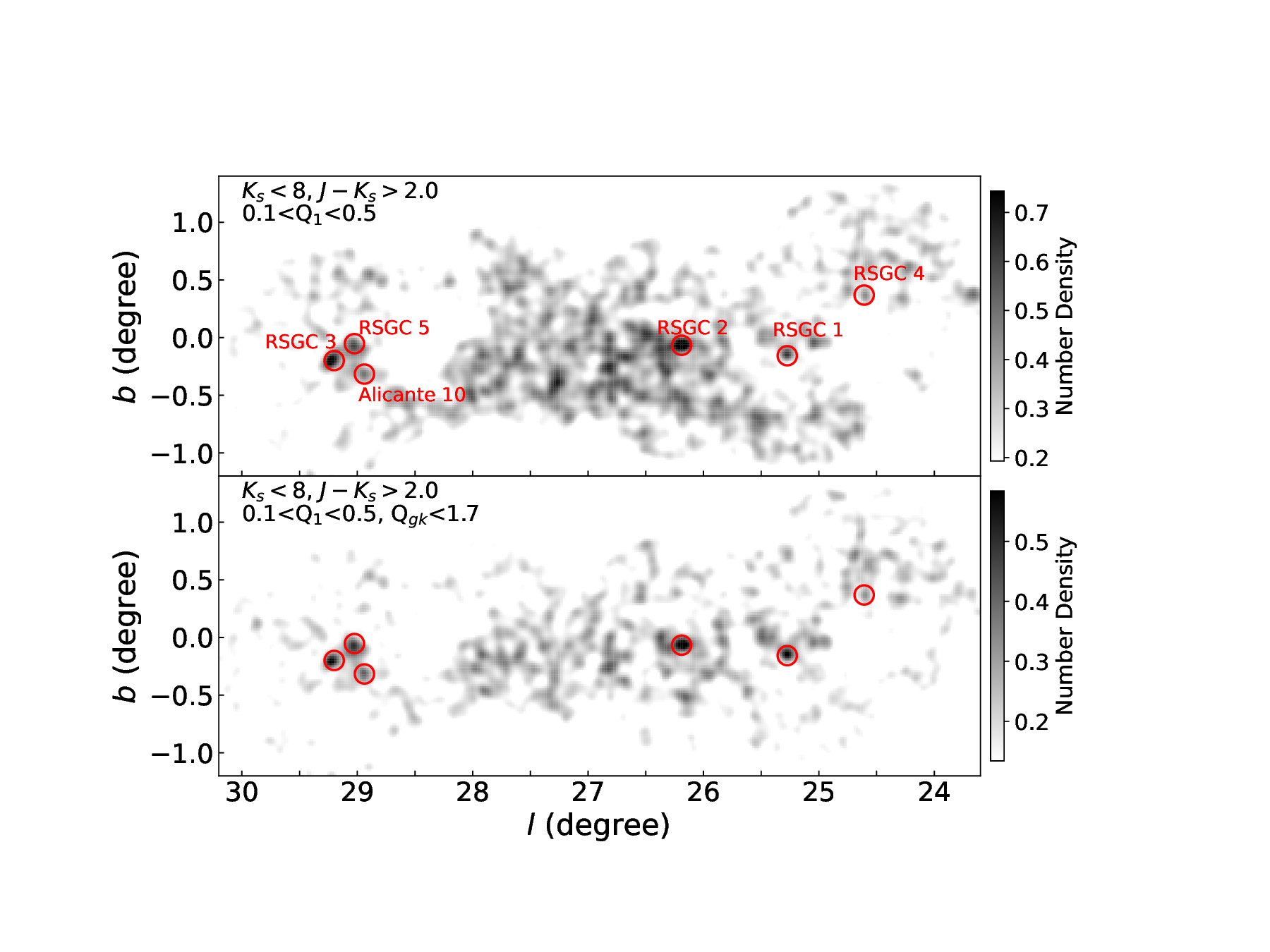}
\caption{The stellar number density maps for the Scutum complex region constructed using different selection criteria. 
The selection criteria are indicated in each panel. The positions of the six RSGCs are indicated by red circles with labels.
The color bars in the each panel show the number density smoothed by the Gaussian filter.
}\label{density}
\end{figure}

Assuming that stars with $Q_{GK_s}>1.7$ were not RSGs and cluster members,
we investigated the overdensities associated with six clusters in the sky.
Figure~\ref{density} shows the stellar density maps for the $\sim6^\circ\times2^\circ$ field of view around six clusters. 
The density map in the upper panel was constructed by selecting the stars that met conditions similar to those (i.e., $K_s<8$, $J-K_s > 2$, and $0.1<Q_1<0.5$) 
established by~\citet{Negu2010}.
Our additional criterion ($Q_{GK_s}<1.7$) was incorporated into the conditions of~\citet{Negu2010}, and the resulting
density map is presented in the bottom panel.
We counted the stars in bins of $1'.2\times1'.2$ widths, and then the number of stars in bins were
smoothed by Gaussian filter with a width equivalent to 10 bins.
The densities from the background to 6$\sigma$ above the level are depicted in both panels.

In the upper panel of Figure~\ref{density}, we found significant overdensities in the RSGC1, RSGC2, and RSGC3 regions,
along with some overdensities in the RSGC5 and Alicante 10 areas. 
However, a dominant density clump was not found in the RSGC4 area, which
contradicts the finding of a clear overdensity by~\citet{Negu2010}.
In the bottom panel, the clusters, except RSGC4, stand out more clearly from the surrounding field when the stars with $Q_{GK_s}>1.7$
are discarded. However the
clumpy structure of RSGC4 appears to be a random fluctuation similar to the background field. 
Therefore, we cautiously suggest that RSGC4 could be a random coincidence of foreground AGBs and RSGs along the line of sight.

\subsection{Binary systems in cluster}
Massive stars with initial masses of $\sim8-30M_\sun$ are frequently found in binary systems~\citep[e.g.,][]{Kobul2007,Sana2013,Dunstall2015,Almeida2017}. 
Interaction with their companion stars during their lifetimes significantly alters their structures and final fates~\citep{Sana2012,Kobul2014}.
The orbital motion of a binary system in a cluster can introduce a large scattering in the radial velocity distribution. 
Therefore, detecting periodic radial velocity variations in our RSG candidates would help constrain the binary system in RSGS4.
However, this is a long-term endeavor requiring long-time observations over a few thousands of days.
 
Most detected RSG binary systems are composed of RSG+B-type companions~\citep[e.g., 108 Galactic RSG+B binary systems;][]{Pantal2020},
as predicted by stellar evolution considerations. The unique spectral features of B-type companions, such as
Balmer absorption lines with additional flux in the UV spectra of B stars or the H$\alpha$ emissions of Be stars, could be detected, as already shown in 
the RSG+B binary spectra obtained by~\citet{Neug2019}.
However, detecting the spectral features of the B-type companion stars in our near-infrared spectra is impossible.
Instead, the effect of the strong interactions between companions can be observed in the spectra. Mass transfer from previously evolved stars (e.g., RSG+WR, although rare) could have altered the atmospheric chemical compositions of the RSGs~\citep[e.g.,][]{Smith1988}. 
Significant mass loss toward the companions may have eliminated the outer envelopes of the RSGs. 
In these processes, the feature of chemical pollution that cannot be explained by a single normal RSG evolution can be detected in some RSG binary systems.
Therefore, we searched for any peculiar spectral features in our high-resolution RSG spectra and compared them with those for typical RSGs.

\begin{figure*}
\includegraphics[width=1.0\textwidth]{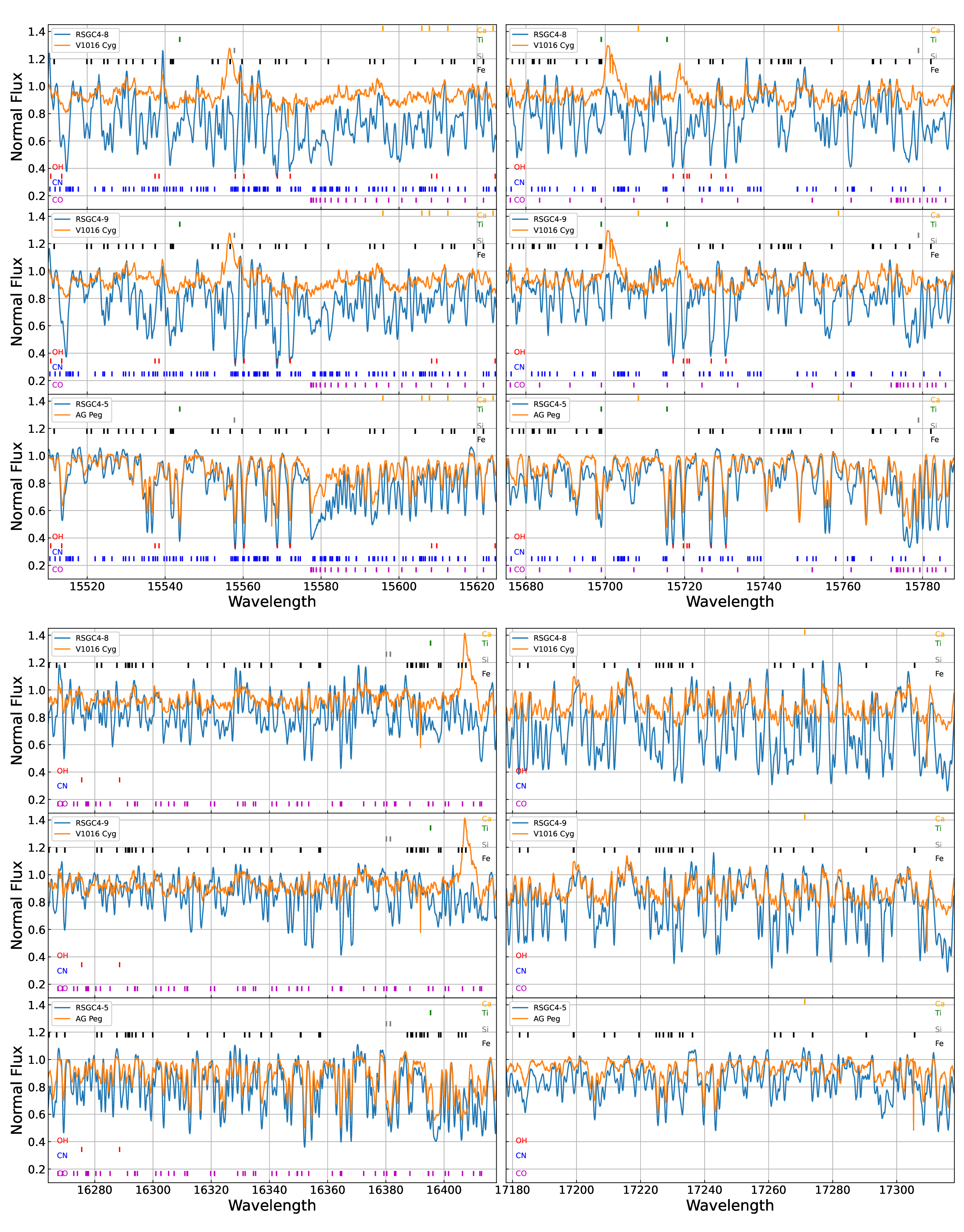}
\caption{Comparison of the $H$-band spectra of RSGC4-8, RSGC4-9 (peculiar RSGs), and RSGC4-5 (normal RSG). The spectra of V1016 Cyg (D-type) and AG Peg  (S-type) symbiotic stars are also described as orange solid lines to show similarities in spectral absorption features. Several atomic (Si, Ca, Ti, and Fe)  and molecule (OH, CN, and CO) absorption lines are indicated by vertical lines of different colors.
}\label{Hcompare}
\end{figure*}

\begin{figure*}
\includegraphics[width=1.0\textwidth]{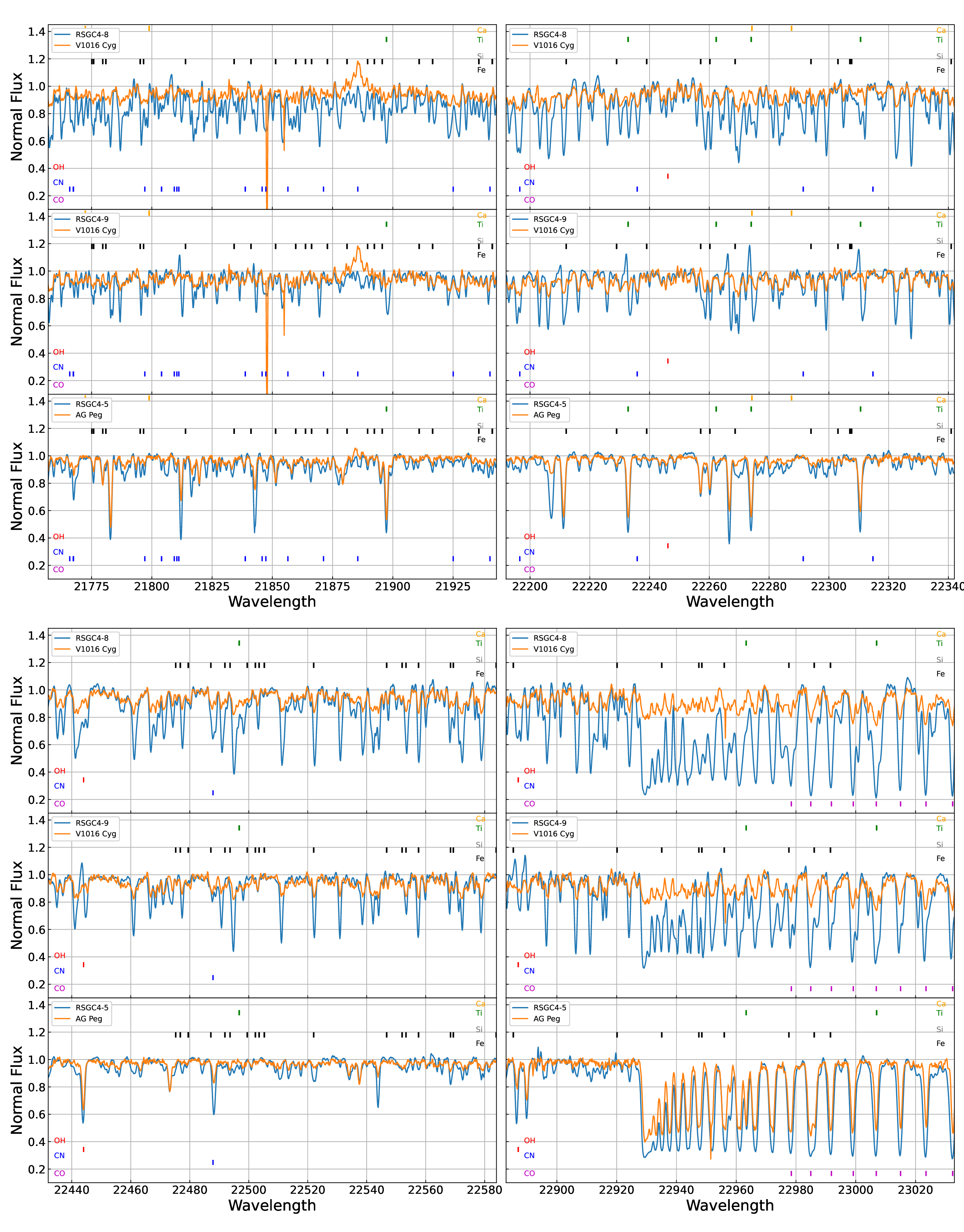}
\caption{Comparison of the $K$-band spectra of RSGC4-8, RSGC4-9, and RSGC4-5. 
}\label{Kcompare}
\end{figure*}

Figure~\ref{Kspec} shows that the spectra of RSGC4-1, 8, 9, and 12 (hereafter referred to as peculiar RSG candidates) display different spectral features from those of the other RSG candidates in RSGC4.
We note that the latter RSG candidate spectra show spectral shapes and absorption features similar to those of the RSG candidates in the other five clusters.
Thus, we consider these RSG candidates to be representative of normal RSG candidates in the Scutum complex region.
The spectra of RSGC4-1, 8, 9, and 12 have a relatively low CO bandhead absorptions compared to those of normal RSG candidates in RSGC4, whereas the
atomic absorptions of RSGC4-8 and RSGC4-9 are much deeper than those of the other normal RSG candidates 
(see Figures~\ref{Kspec}, ~\ref{Hcompare} and ~\ref{Kcompare}).
Although there are slight differences in the atomic absorption strengths, the spectra of RSGC4-8 and RSGC4-9 closely resemble each other.
The spectra of RSGC4-1 and 12 simultaneously show spectral absorption features of both the normal RSG candidates and RSGC4-8 and 9.

We compared the spectra of peculiar RSG candidates with single normal RSG model spectra.
The several atomic abundances in model spectra were adjusted to mimic the observed spectra. However, we could not simulate any theoretical spectra that closely resembled the observed absorptions of the peculiar RSG candidates.
Notably, the spectra of normal RSG candidates were well represented by single RSG models, although the atomic and molecular linelists~\citep{Afsar2016} could not perfectly describe the RSGs.
Thus, there seem to be some unknown physical properties of peculiar RSG candidates that cannot be explained by single RSG models.

We searched the several spectral libraries and the relevant literature to find spectra similar to those of peculiar RSG candidates. 
Eventually, we found a hint that their spectra were similar to the spectra of D-type symbiotic stars (e.g., V1016 Cyg).
Symbiotic stars are interacting binary systems consisting of an evolved cool giant and a compact companion, typically an AGB star and a white dwarf.
Fortunately, several symbiotic stars were also observed using IGRINS and are available in the Raw \& Reduced IGRINS Spectral Archive (RRISA)\footnote{~\url{https://igrinscontact.github.io/}}. 
Thus, we could directly compare our spectra with those of symbiotic stars. 

Figures~\ref{Hcompare} and~\ref{Kcompare} show the spectra comparisons of RSGC4-8 and 9 as peculiar RSG candidates and 5 as a normal
RSG candidate in the $H$- and $K$-bands, respectively. The spectra of the AG Peg and V1016 Cyg symbiotic stars are also included as orange solid lines.
AG Peg is a normal S-type symbiotic star, and its spectrum is characterized by a stellar continuum and the absorptions of normal late-type giants.
Therefore, the spectrum of AG Peg do not differ significantly from the spectrum of RSGC4-5 (i.e., normal RSG candidate), as shown in Figures~\ref{Hcompare} and~\ref{Kcompare}.

In contrast, V1016 Cyg is a famous D-type symbiotic star comprising a Mira-AGB and a white dwarf. 
This star shows an infrared (IR) excess from the dust produced by the evolved Mira-AGB variable.
Interestingly, in Figures~\ref{Hcompare} and~\ref{Kcompare}, the spectrum of V1016 Cyg not only shows entirely different absorption features from those of AG Peg and RSGC4-5 but also closely 
traces very well the peculiar absorption features of RSGC4-8 and 9.
The only difference between V1016 Cyg and the peculiar RSG candidates is the presence of strong emission lines in V1016 Cyg.
It is not clear why our peculiar RSG candidates show no emission lines, but if there was no shell-burning in the accretion region around the hot component,
weak emission or no emission would be possible~\citep[e.g., SU Lyncis;][]{Mukai2016}. 
Considering this point, our peculiar RSG candidates in RSGC4 are not precisely D-type symbiotic stars but seem to exhibit similar or corresponding properties to D-type symbiotic stars.  

\begin{figure*}
\includegraphics[width=\textwidth]{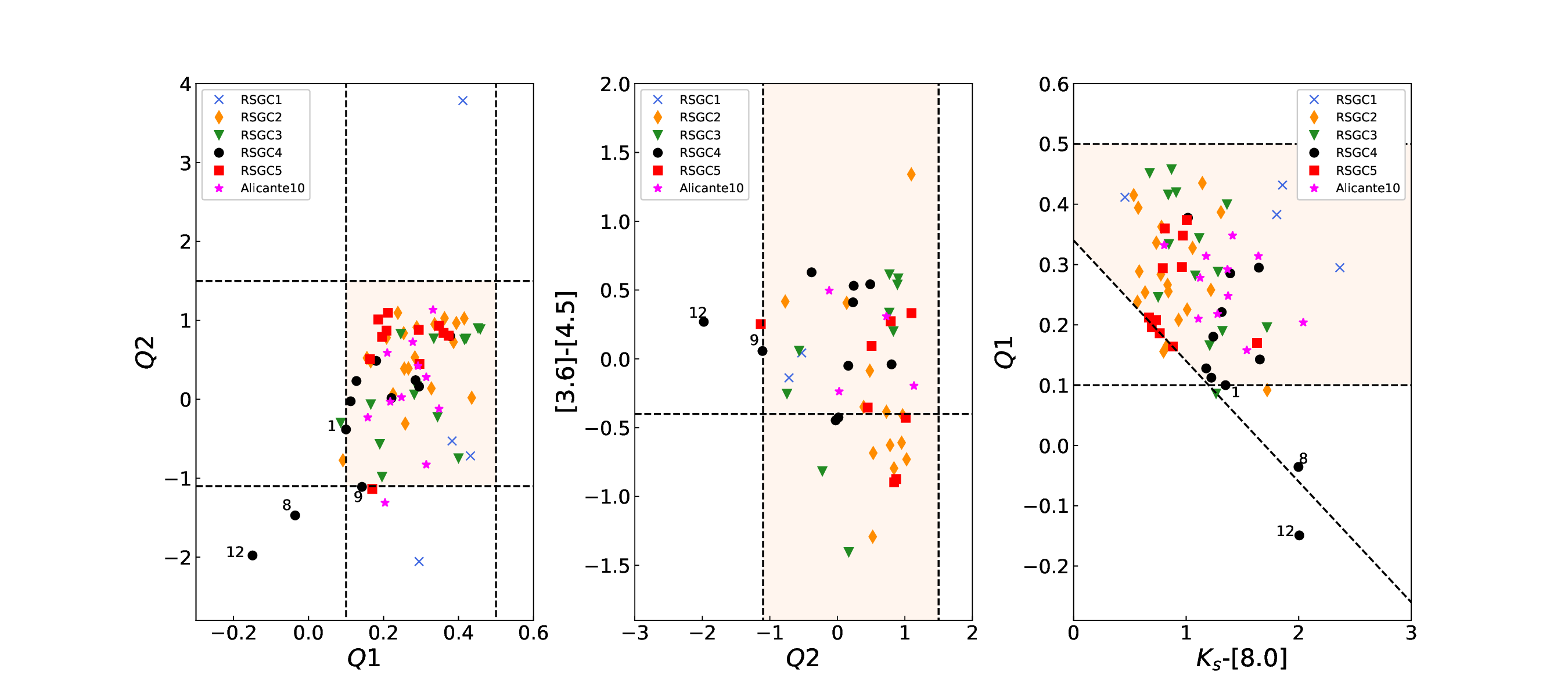}
\caption{The distributions of RSG candidates in six RSGCs in $(Q1, Q2)$, $(Q2, [3.6]-[4.5])$, and $(K_s-[8.0], Q1)$ CCDs. The dashed lines indicate the criteria of RSG established by~\citet{Messi2012},
and the shaded region in each panel is where most Galactic RSGs are distributed. The peculiar RSG candidates with ID numbers in RSGC4 are located in the region where IR excesses are expected.
}\label{CCD}
\end{figure*} 
One of the most prominent properties of D-type symbiotic stars is the IR excess resulting from dust produced by an evolved Mira-AGB variable.
~\citet{Messi2012} conducted a photometric classification of Galactic evolved stars, analyzing their distributions in near- and mid-infrared 
color-color diagrams using the 2MASS and Galactic Legacy Infrared Midplane Extraordinaire (GLIMPSE) catalogs.
Using the reddening-free $Q1$ and $Q2$\footnote{$Q2=(J-K_s)-2.69\times(K_s-[8.0])$} parameters, GLIMPSE $[3.6]-[4.5]$, and $K_s-[8.0]$ colors, 
they assigned specific color regions for objects with IR excess and RSGs. 
We also examined the same CCDs for the RSG candidates in our RSGCs to assess their IR excess properties.
The magnitudes in the mid-infrared region of GLIMPSE are provided in Table~\ref{photo}.
 
Figure~\ref{CCD} shows the distributions of RSG candidates in the six RSGCs in the (Q1, Q2), (Q2, $[3.6]-[4.5]$), and ($K_s-[8.0]$, Q1) CCDs. 
As determined by~\citet{Messi2012}, the shaded region in each CCD indicates
the area where most Galactic RSGs are distributed; $0.1<Q1<0.5$, $-1.1<Q2<1.5$, and $Q1>-0.2\times(K_s-[8.0])+0.34$.
Apparently, most of RSG candidates in the six RSGCs are situated inside the shaded area of normal RSGs, while three or four peculiar RSG candidates
in RSGC4 are located in the region of IR excess: $Q1<0.1$, $Q2<-1.1$~\citep{Messi2012} and $K_s-[8.0]>0.5$~\citep{Vanloon2003}. 
Therefore, these peculiar RSG candidates are likely to be the stars with circumstellar materials due to their high mass loss.

The mass-loss rates of RSGs range from $\sim10^{-7}$ to $\sim10^{-4} M_\sun$yr$^{-1}$~\citep[e.g.,][]{Josselin2000,Verho2009}, 
which are comparable to Mira-AGBs~\citep[from $\sim10^{-7}$ to $\sim10^{-4} M_\sun$yr$^{-1}$;][]{Habing2003} but not 
maser sources~\citep[from $\sim10^{-5}$ to $\sim10^{-3} M_\sun$yr$^{-1}$;][]{Ortiz2002}.
We note that Mira-AGBs and SiO or OH/IR maser stars also have an IR excess and can be located in the IR-excess areas on the CCDs.
However, as mentioned earlier, we did not detect Mira-AGB variability. No maser sources in this field were reported by~\citet{Negu2010}.
In this respect, the peculiar RSG candidates in RSGC4 do not appear to be Mira-AGBs or maser sources. 
They seem to be the binary systems containing AGB or RSG stars
surrounded by dust, resulting from the significant mass loss by their companions.

We obtained additional IGRINS spectra for some RSG candidates in RSGC4 in 2021 to examine radial velocity changes. The measured velocities are
indicated in Table~\ref{kinematics}. 
The radial velocities of seven RSG candidates (RSGC4-1, 2, 3, 4, 5, 6, and 7) in RSGC4, observed in 2019 by~\citet{Asad2023}, are also
included in Table~\ref{kinematics}.
Most of the RSG candidates observed in 2021 and by~\citet{Asad2023} show slight velocity changes of 2--3 km $\mathrm{s^{-1}}$,
which is acceptable considering the vigorous atmospheric activity of RSGs. 

Interestingly, however, two peculiar RSG candidates with D-type absorption features in our study
(i.e., RSGC4-8 and 9) show different velocity changes. RSGC4-9 show a large velocity variation of approximately $20$ km $\mathrm{s^{-1}}$.
Typical radial velocity variations due to binary motions are between 20 km s$^{-1}$ and 30 km s$^{-1}$~\citep[e.g.,][]{Wright1977,Eaton2007,Patrick2019}.
On the other hand, the velocity variation of RSGC4-8 is about 5 km s$^{-1}$ and changes rapidly within a month. This phenomenon contradicts the long-term
variation expected of an RSG binary system. 
Another peculiar RSG candidate, RSGC4-1, shows a velocity change of about 5--8 km s$^{-1}$. 
Thus, if these three peculiar RSG candidates are real binary systems, it is highly likely that these systems contain high-mass AGBs, consistent with the 
foreground AGB contamination results described in the previous section.
It is also interesting that RSGC4-3 exhibits the second largest velocity change (15--17 km s$^{-1}$) despite its spectrum and
photometric colors appearing to be those of a normal RSG candidate.
To conclude on the binarity of this star, additional radial velocity monitoring is necessary.

Regardless of whether our peculiar RSG candidates are indeed AGBs or genuine RSGs, we have concluded that only five stars (RSGC4-1, 3, 8, 9, and 12) 
among the 12 RSG candidates in RSGC4 have the potential features of binary systems in terms of spectra, photometric colors, and variations of radial velocity. 
Among these, three (RSGC4-3, 8, and 9) are confirmed binary systems. 
When we rejected these five RSG candidates in the cluster, there still remained a large scattering of 46 km s$^{-1}$ in the velocity distribution.
Although additional multi-epoch observations for other RSG candidates are needed, 
it is apparent that the effect of binary motion alone cannot fully explain the large scattering in the radial velocity distribution and the offset 
from the mean velocity of the RSG candidates in the RSGCs.

\subsection{Unbounded stars in the cluster}
\begin{figure*}
\includegraphics[width=\textwidth]{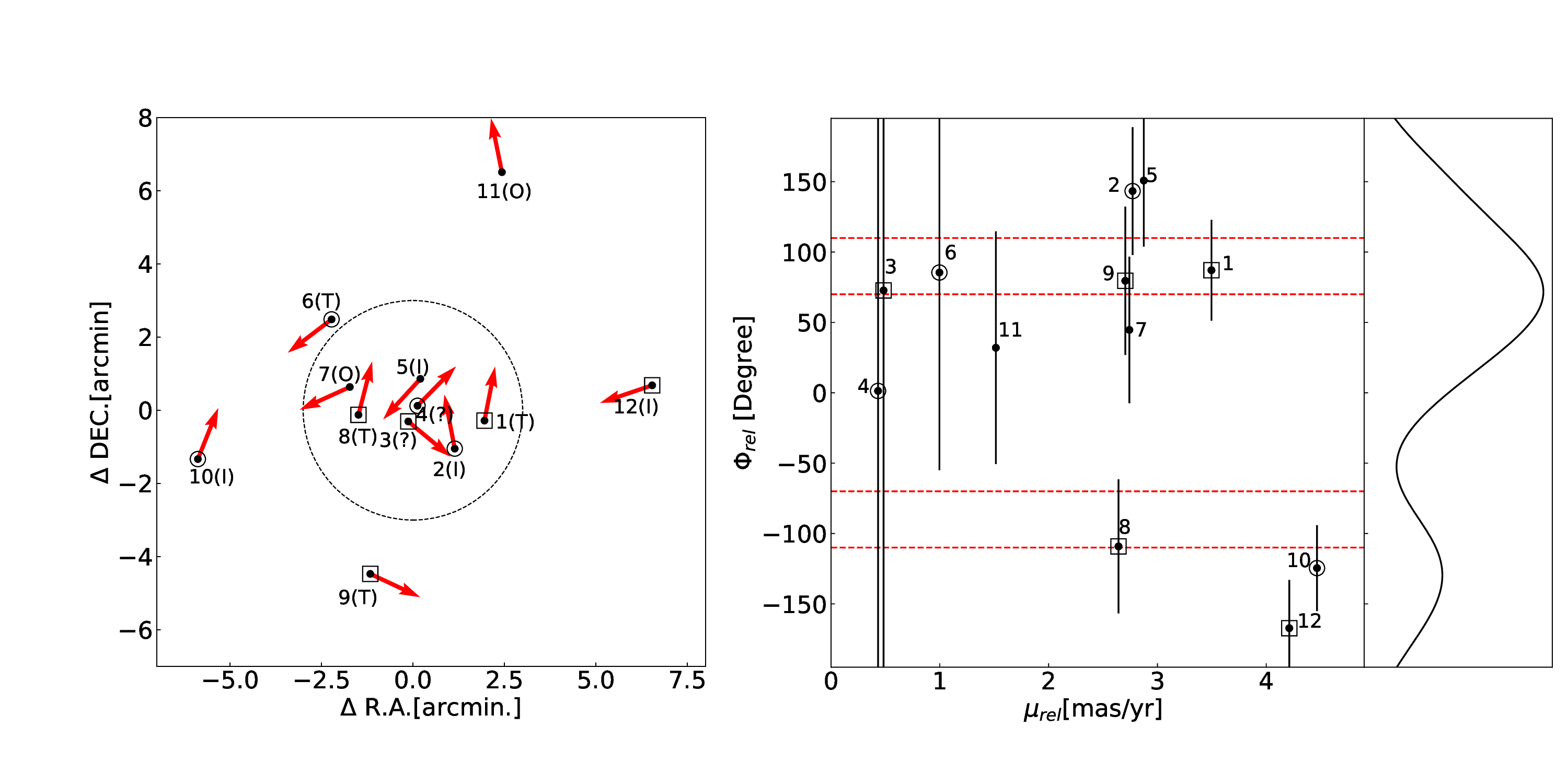}
\caption{Left: Relative spatial distribution of RSG candidates in RSGC4 from the cluster center. The relative PM vectors are also indicated by arrows. The dotted circle indicates the region within $3'$ of the cluster center. The ID numbers of candidates are indicated, and the approximate motions of candidates based on $\Phi_{\mathrm{rel}}$ are indicated in parentheses (see text in Section 4.3).
Right: Distribution of 
$\Phi_{\mathrm{rel}}$ as a function of $\mu_{\mathrm{rel}}$ for RSG candidates in RSGC4. The potential binary stars and the stars with RUWE $<$ 1.4 are 
indicated by open squares and open circles, respectively. 
The red dashed lines indicate the angle criteria ($\pm70\arcdeg$ and $\pm110\arcdeg$) used to determine the approximate motions of RSG candidates.
The KDE plot of $\Phi_{\mathrm{rel}}$ is shown in the right-hand panel.
}\label{motion}
\end{figure*}

A large offset velocity of RSG candidates in RSGC4 of more than 30 km s$^{-1}$ compared to the mean velocity of RSGs in the Scutum complex reminds us of
potential runaway or walkaway stars~\citep{Blaauw1956,Blaauw1961,Gies1986,Dray2005,Eldridge2011,Boubert2018}.
Given that a significant fraction (10\%--50\%) of OB main-sequence stars are runaway stars~\citep{Gies1986}, and most of them undergo
the RSG evolutionary phases, we expect that relatively numerous RSGs would be found as runaway or walkaway stars.
However, very few runaway RSGs with peculiar velocities have been found so far in the Milky Way: Betelgeuse at 56 km s$^{-1}$~\citep{Nori1997},
$\mu$ Cep at 22 km s$^{-1}$~\citep{Cox2012}, and IRC-10414 at 70 km s$^{-1}$~\citep{Gvara2014}.
 
The RSG candidates in RSGC4 are probably runaway stars with low velocities~\citep{Gvara2012}, or simply
escaping stars from the cluster.
Since one-dimensional radial velocity alone cannot precisely describe the motion of stars, we examined the proper motion (PM) of RSG candidates in RSGC4
from $Gaia$ DR3 data. 
We identified 12 RSG candidates in RSGC4 from $Gaia$ DR3 data, and the proper motions are 
presented in Table~\ref{kinematics}. If the several RSG candidates in RSGC4 are real runaway or walkaway stars, they may
be placed in the outer region of the cluster, and the moving direction of the stars could show outward movement from the center of the cluster.

The relative spatial distributions of the RSG candidates in RSGC4 from the center of the cluster are plotted in the left panel
of Figure~\ref{motion}. 
The relative PM vectors of stars to the median PM ($\mu_{\alpha,\mathrm{med}}$cos $\delta_\mathrm{med}=-1.334$ mas yr$^{-1}$ and $\mu_{\delta,\mathrm{med}}=-5.417$ mas yr$^{-1}$)
of RSG candidates are shown as red arrows in the left panel of Figure~\ref{motion}. We have indicated the direction of the relative PM vectors, and the 
length of the arrows does not represent the magnitude of the velocity. 
The center of the cluster was determined by the median value of the sky positions of the RSG candidates.

In Figure~\ref{motion}, five RSG candidates (RSGC4-6, 9, 10, 11, and 12) appear to be located more than $3'$ from the cluster center, while
seven RSG candidates are closely grouped together within $3'$ of the center, forming a clump.
We note that the potential binary stars in RSGC4 (open squares) have large Renormalised Unit Weight Error (RUWE) values ($>1.4$) of $Gaia$ DR3 data.
The relative PM vectors seem to show inward motions, contradicting the expected outward motion of runaway or walkaway stars.

To investigate inward motion in detail, we calculated the orientation angles of relative PM vectors, ($\Phi_{\mathrm{rel}}$), between the radial vector
of a given star from the center of the cluster and its relative PM vector. 
We described this angle as a function of the amplitude of the relative PM vectors in the right panel of Figure~\ref{motion}.
The amplitude of the relative PM vectors was calculated using the following equation:
\begin{equation}
\mu_{\mathrm{rel}} = \sqrt{(\mu_{\alpha}\cos\delta-\mu_{\alpha,\mathrm{med}}\cos\delta_{\mathrm{med}})^2+(\mu_{\delta}-\mu_{\delta,\mathrm{med}})^2}.
\end{equation}
The value of $\Phi_{\mathrm{rel}}$ describes the overall expansion or contraction motion of stars in the cluster, as demonstrated by~\citet{Lim2019};
stars with $\Phi_{\mathrm{rel}}=0\arcdeg$ have expansion motion, while those with $\pm180\arcdeg$ have contraction motion.

In the right panel of Figure~\ref{motion}, it is difficult to decide whether the stars have expansion or contraction motions due to the large errors.
However, the RSG candidates in RSGC4 seem to be distributed around $-150\arcdeg$ and 80$\arcdeg$ of $\Phi_{\mathrm{rel}}$ in Kernel Density Estimate (KDE) plot.
The bimodal distribution in $\Phi_{\mathrm{rel}}$ suggests marginal inward motions of RSG candidates in RSGC4.
Based on $\Phi_{\mathrm{rel}}$, we determined approximate motions of RSG candidates as following criteria:
\begin{itemize}
\item $|\Phi_{\mathrm{rel}}|>110\arcdeg$ : inward motion (`I')
\item $70\arcdeg<|\Phi_{\mathrm{rel}}|<110\arcdeg$ : tangential motion (`T') 
\item $|\Phi_{\mathrm{rel}}|<70\arcdeg$ : outward motion (`O')
\item Undecided due to large error : (`?')
\end{itemize}
Determined motions are indicated in the left panel of Figure~\ref{motion}.

According to $\Phi_{\mathrm{rel}}$, there are 4 stars exhibiting inward motions, 4 stars showing tangential motions, and 2 stars displaying outward motions.
The bimodal distribution of $\Phi_{\mathrm{rel}}$ in the KDE plot represents the stars with tangential and inward motions. 
We do not find any spatial dependence among stars with tangential and inward motions.
The stars concentrated within $3'$ of the center, as well as those outside of it, exhibit various motion patterns.
Among the four RSG candidates with peculiar absorption lines, only one star (RSGC4-12) exhibits
inward motion, while RSGC4-1, 8, and 9 show tangential motions to the radial vector from the center of the cluster.
We find that only two stars (RSGC4-7 and 11) among 12 RSG candidates in RSGC4 are likely to show outward motion. 
We also investigated the $\Phi_{\mathrm{rel}}$ distributions of RSG candidates in other clusters; notably, we found no distinct expansion or contraction motions of 
RSG candidates in the other clusters.

The marginal inward and tangential motions of the stars indicates that 
the RSG candidates in RSGC4 are neither the runaway nor walkaway stars from the cluster.
Considering the young age of $\sim$20 Myr for this cluster~\citep{Negu2010}, the inward motions and concentrations of some RSG candidates could indicate mass segregation.
The absence of main-sequence stars~\citep[e.g.][]{Froe2013} and the more spatially extended feature of this cluster
compared to others may support the idea of mass segregation in this cluster.
However, in this case, the reason for the large spread in the radial velocity of the bounded RSG candidates resulting from mass segregation remains unclear.

\subsection{Environmental effects}
\begin{figure*}
\plottwo{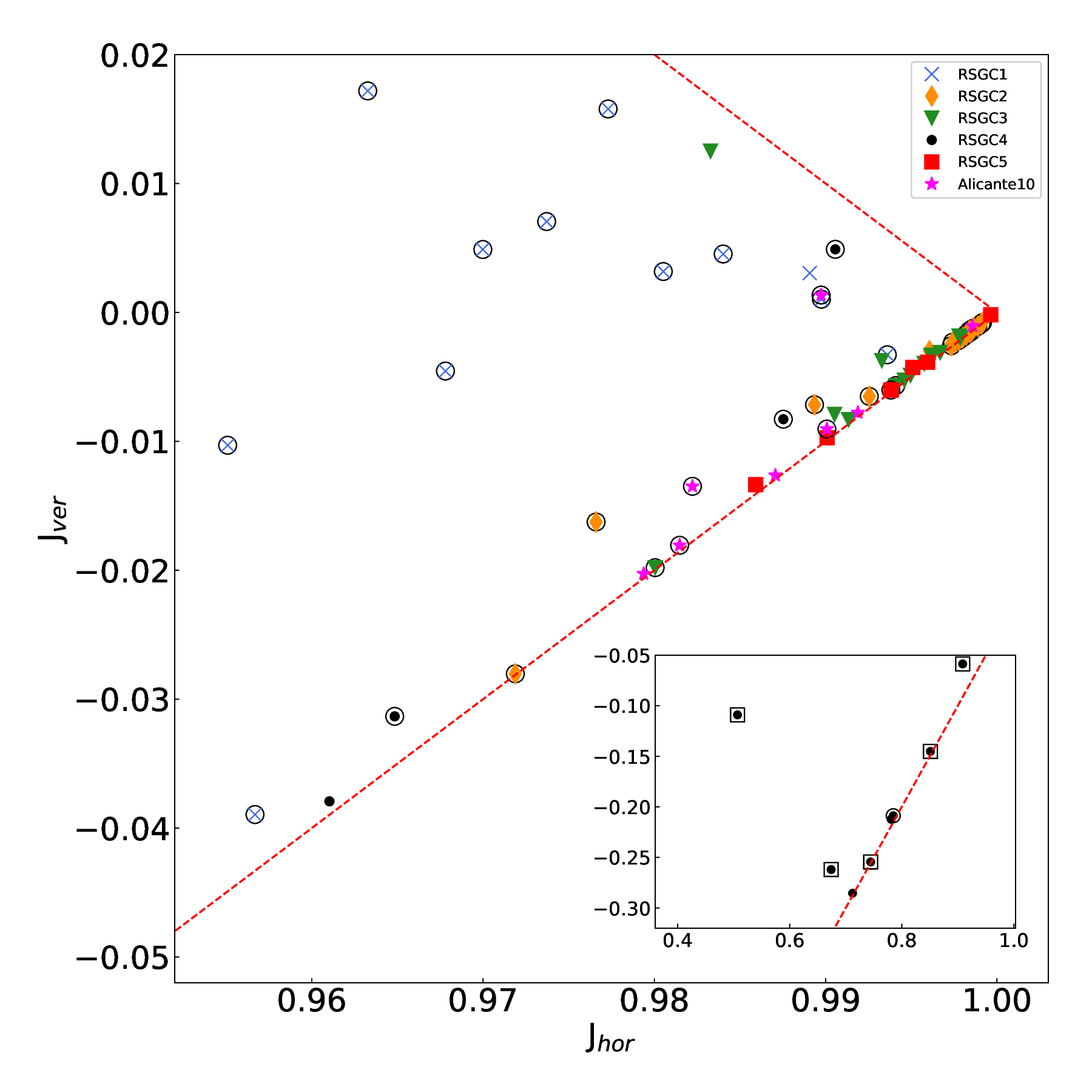}{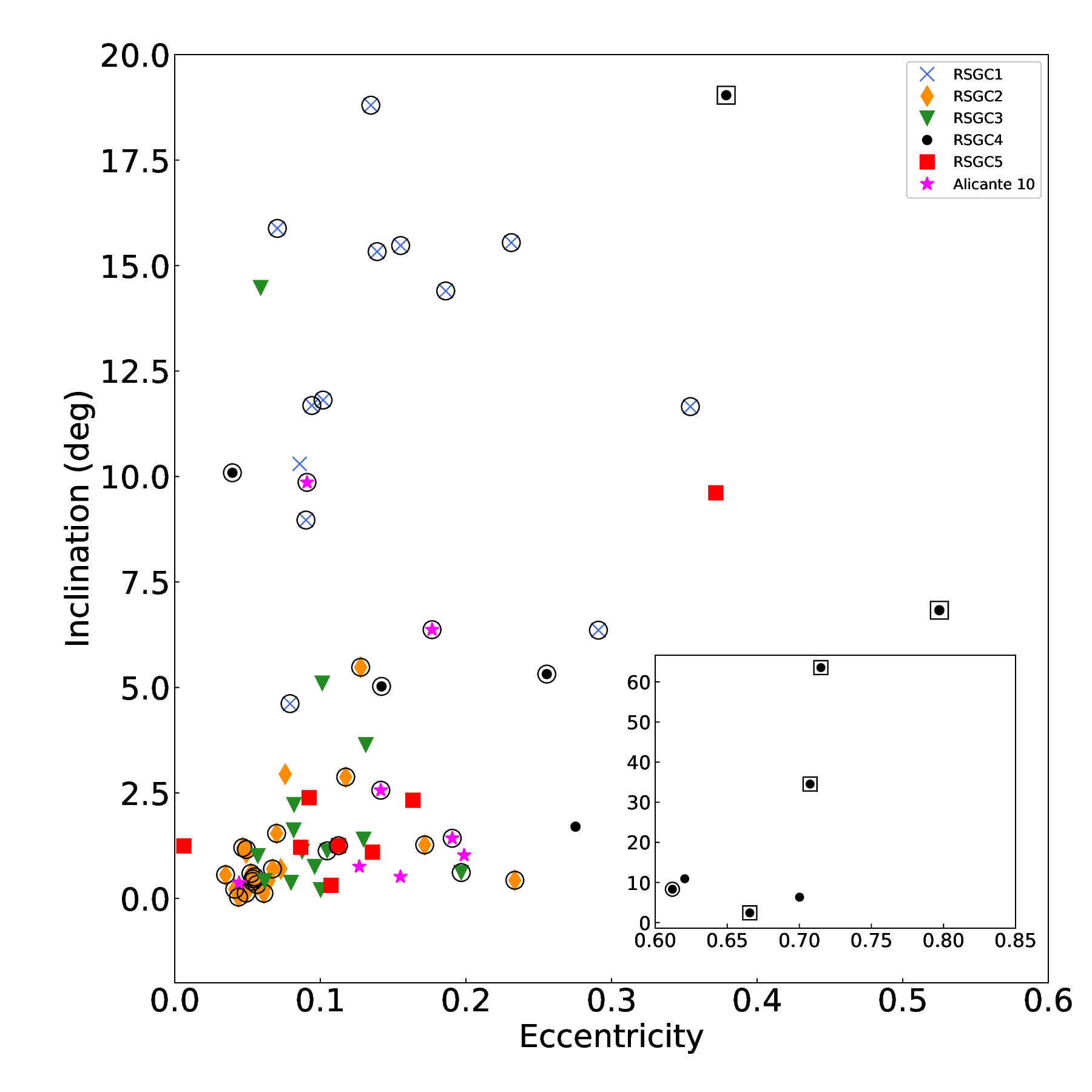}
\caption{Left: Action space map of normalized ($J_\mathrm{hor}, J_\mathrm{ver}$) for the RSG candidates in the six RSGCs. The RSG candidates in RSGC4 with low $J_\mathrm{hor}$ and $J_\mathrm{ver}$ are indicated in the subplot of the figure. The stars with RUWE $<$ 1.4 are indicated by open circles. 
The potential binary stars in RSGC4 are indicated by open squares. The red dashed line shows the limits of the action space map.
Right: Map of inclination against eccentricity for the RSG candidates in the six RSGCs. 
The RSG candidates in RSGC4 with high eccentricity are shown in the subplot.}
\label{action}
\end{figure*}
The close proximity of RSGCs to the junction where the spiral arms meet the Galactic bar naturally 
leads us to consider the environmental effects of the unusual radial velocities of the RSG candidates in RSGC4. 
The spiral arms or bars of the galaxy are places where inflowing gaseous material and stars from the outer disk are easily accumulated.
Thus, cloud-cloud collisions are expected in these regions. These collisions inevitably lead to strong star formation~\citep{Loren1976,Tan2000,Sheth2005,Inoue2013} and massive star formation~\citep{Anath2010,Motte2014}.

Particularly at the Lindblad resonance radii, where the bar intersects with the spiral arm, stars and gas experience shocks and loss angular momentum~\citep{Combes1985}. Consequently, the material migrates toward the central regions, yielding a ring-like structure~\citep[e.g.,][]{Schwarz1981,Shlosman1990,Athan1992} characterized by intense star formation and 
nuclear starburst~\citep[e.g.,][]{Knapen1995,Wong2000,Mazzuca2008}. Indeed, the presence of numerous massive stars~\citep{Lopez1999}, RSGC clusters~\citep{Figer2006,Davies2007,Clark2009}, and stellar and H II regions with ring-like structures~\citep{Bertelli1995,Davies2009} at the outer Lindblad resonance of the Galactic bar indicates recent intense starburst activity in this region. 

In this respect, the tip of the bar is a dynamically complex structure, and the young stars and massive star clusters formed in this region are 
significantly influenced by their surroundings. For example, the tidal force from the bar potential or shocks resulting from encounters with dense giant molecular clouds can have a strong impact on the morphology and dynamic evolution of these clusters, sometimes leading to their complete dissolution~\citep{Spitzer1958,Gnedin1999,Gieles2006}. 

Recently, ~\citet{Whitmore2023} provided new insights into the formation of young massive star clusters in the bar region using JWST 
observations of NGC 1365.
They reported that the collision of material streamers with gas in the bar and overshooting stars from the opposite bar lane play crucial roles in star cluster formation. On the other side, it is also known that the formation of spiral arms and bars in the Milky Way is influenced by the merging and accretion of gas-rich dwarf galaxies~\citep{Purcell2011}. Several theoretical results~\citep{Schweizer1987,Ashman1992} and observational evidence~\citep{Vanden1971,Arp1985,Larsen2006,Roche2015} suggest that such mergers lead to the formation of massive star clusters.
Therefore, the peculiar radial velocity properties of the RSG candidates in RSGC4 could be the result of secular evolution influenced by environmental factors. 

Based on the radial velocities, distances to the clusters from literature (see Section 4.1), and proper motions of $Gaia$ DR3 data,
we derived the actions and orbital properties of RSG candidates in the six RSGCs using \Agama galaxy modeling architecture~\citep{Vasiliev2019} to
investigate the dynamical properties affected by their surroundings. 
We utilized the potential of~\citet{McMillan2017}  and the St\"ackel fudge approach~\citep{Binney2012}. The map of normalized horizontal action $J_\mathrm{hor}=J_\mathrm{\phi}/J_\mathrm{tot}$ and vertical action $J_\mathrm{ver}=(J_\mathrm{z}-J_\mathrm{R})/J_\mathrm{tot}$, where $J_\mathrm{tot}$ is the sum of absolute values of $J_\mathrm{R}, J_\mathrm{\phi}$ and $J_\mathrm{z}$~\citep[see e.g.,][]{Myeong2019}, is shown in the left panel of Figure~\ref{action}.

The left panel of Figure~\ref{action} shows that most of the RSG candidates, except in RSGC1 and RSGC4, have disk-like motions with a horizontal action of $J_\mathrm{hor}\sim1.0$ and a vertical action of $J_\mathrm{ver}\sim0.0$, consistent with the expected behavior of stars in the disk. 
In contrast, the RSG candidates in RSGC4 show a large scattering in both the $J_\mathrm{hor}$ and $J_\mathrm{ver}$ distributions. 
Interestingly, they have relatively low values for both $J_\mathrm{hor}$ and $J_\mathrm{ver}$ compared to the other RSG candidates.
Other interesting features are unexpected scattering and the relatively large vertical actions of the RSG candidates in RSGC1.
These dynamical properties are also clearly shown in the eccentricity and inclination 
in the right panel of Figure~\ref{action}. Half of RSG candidates in RSGC4 show eccentricities greater than 0.6, and the majority have inclinations of more 
than $5^\circ$. 
Most RSG candidates in RSGC1 have a large inclination of more than $10^\circ$ compared to the other RSG candidates.
Although several RSG candidates have a RUWE $>$1.4 and more accurate astrometric solutions are necessary, it is important to note that the current action values,
eccentricities, and inclinations of the RSG candidates possess a certain level of reliability. For example, as expected, a large number of RSG candidates 
with RUWE$>$1.4 in other clusters still show disk-like motions, and most RSG candidates in RSGC1 have RUWE values less than 1.4. 
We confirmed that the dynamical properties of each cluster, indicated by the action values, remained relatively stable even when the distance to each cluster or proper motion was changed, with offset values applied.

The vertical motion of RSGC1 and the high eccentricity of RSGC4 might indicate dynamical evolution resulting from the environmental effects that the clusters
experienced during their formation in single-star forming event. Encounters with dense and giant molecular clouds in the disk could influence the dynamical evolution of clusters, increasing their vertical motion. Tidal forces from bar potential or local gaseous substructures~\citep[e.g.,][]{Elmegreen2010} in the Scutum complex region can contribute to the disruption of clusters.

~\citet{Park2020}, in their $\sim2$ Myr timespan simulation, have shown that a strong tidal field can dissolve a single star-forming cloud, 
resulting in several bound star clusters that contain an overabundance of massive stars of $20~\mathrm{M}_\sun$.  
Notably, the distribution of star clusters in their simulation (see the rightmost panel of their Figure 5) closely resembles the distribution of RSGCs 
in the sky shown in Figure~\ref{density}. The small, sparse group of massive stars labeled `2b' in their Figure 5 could be a candidate for RSGC4.
However, it remains unclear how these environmental factors could selectively affect only two out of the six clusters at almost the same distance, even leading to different dynamic evolutionary properties.
To get a better understanding, 
properties of stars in clusters, such as the velocities of massive stars in clusters, are necessary from additional simulation spanning 20 Myr.
Without a detailed comparison with simulation results, it is more likely that for the RSG candidates in RSGC4, 
stars moving toward the bar tip along the bar lane overshoot the bar tip, and then overlap with the stars traveling along the arms or gas streamers. 
In this scenario, the stars moving along the bar lane appear to be moving toward us, while the stars traveling along the arm seem
to be moving away from us.
This interpretation is supported by the discovery of massive stars with ages of 10--20 Myr in several overshoot regions and the NE bridge feature
of NGC 1365 using JWST~\citep{Whitmore2023}.

Additionally, the distribution of a significant number of RSGCs in the narrow Scutum complex region raises the possibility that RSGC4 might
 be a binary cluster or an interacting cluster.
Previous studies by~\citet{Rozh1976} and~\citet{Subra1995} suggested that 8--20\% of open clusters born in star complex regions could be binary clusters.
Resonant trapping in the Scutum complex region can potentially lead to the formation of interacting or merging star clusters~\citep[][]{Dela2009}.
Therefore, the distinct radial velocities and action values observed for the RSG candidates in RSGC4 could be attributed to such interactions.

In binary cluster scenario, bimodalities or clumpy structures~\citep{Kuhn2014,Darma2019} in the spatial positions of cluster member stars can be expected.
Although cluster pairs are likely to share common kinematics, detecting slightly different motions among member stars belonging to different clusters is plausible, considering their young ages.
Despite these expectations, we could not find apparent bimodalities in the spatial positions and motions, including radial velocities, proper motions, and even action values, of the RSG candidates in RSGC4.
To confirm the binary cluster scenarios for RSGC4, conducting additional statistical tests on the spatial and motion properties of main-sequence members within the cluster would be helpful.

\section{Summary and conclusion}
In this study, we obtained high-resolution ($R\sim45,000$) near-infrared spectra for 60 RSG candidates in six RSGCs in the Scutum complex region, where 
the Scutum-Crux arm intersects the tip of the Galactic bar, using IGRINS at the Gemini South telescope. 
From their spectra, we find that the RSG candidates in RSGC4 share no common radial velocities, unlike the consistent radial
velocities found in the other cluster member stars.
The RSG candidates in RSGC4 show significant scattering in the radial velocity distribution and an offset in the mean velocity compared with the mean radial velocity of the RSG candidates in the other RSGCs. We have discussed four explanations (AGB contamination, binary systems, unbounded stars, and environmental effects) to interpret the unusual radial velocities observed in the RGS candidates of the RSGC4.

AGB contamination was checked by investigating the photometric variations and the new 
reddening-free parameter, $Q_{GK_s}$. 
We find no significant differences between the magnitudes of 2MASS and DENIS. 
Definite variability similar to that of AGB stars was not detected in WISE/NEOWISE data due to the saturation of RSG candidates. 
However, a lower limit of the amplitudes of variation, approximately 1.0 mag in both W1 and W2, was just detected.
Interestingly, we find that most RSG candidates in RSGC4 have $Q_{GK_s}$ values greater than 1.7, which
is the limit that stars in the CHeB evolutionary phase cannot reach. Stars in the early AGB or thermal pulsing AGB phases can exceed this limit and have large values of $Q_{GK_s}$. Indeed, we find that most RSG candidates in the other RSGCs have $Q_{GK_s}$ values below this limit. 
Therefore, it is possible that several RSG candidates in RSGC4 could be massive AGB stars, although we cannot rule out the presence of RSGs surrounded by dust.

We find that the spectra of RSGC4-1, 8, 9, and 12 have different absorption features from those of the other stars in the clusters. 
The peculiar absorptions in their spectra cannot be explained by single normal RSG model spectra, suggesting the possibility that 
these stars are binary systems.
Indeed, the spectra of RSGC4-8 and 9 are very similar to those of D-type symbiotic stars (AGB + white dwarf), except for the absence of emission lines. 
Furthermore, these two stars exhibit large or fast radial velocity variations, providing further evidence of their binary nature. 
In several color-color diagrams, we also find that four RSG candidates with peculiar absorption features have IR excesses similar to D-type symbiotic stars. We also discover radial velocity changes in RSGC4-1 and 3 through multi-epoch observations.
However, even if we excluded these five binary systems from the 12 RSG candidates in RSGC4, significant scattering remained in the velocity distribution.
Therefore, the binary systems cannot fully explain the peculiar radial velocities found in the RSGC4.

We investigated the relative spatial distributions, relative proper motions, and their orientation angles ($\Phi_{\mathrm{rel}}$) of the RSG candidates in RSGC4 to determine whether there were runaway or walkway stars in the cluster.
Despite the limited number of RSG candidates, 7 of 12 stars are located within $3'$ of the center of the cluster.
Slight bimodal distributions in the orientation angles of the relative proper motions indicate 
marginal inward and tangential motions of the RSG candidates in RSGC4.
We did not observe any spatial dependence among stars with tangential and inward motions.
Therefore, the RSG candidates in RSGC4 are neither the runaway nor walkway stars influenced by supernovae events.
 
The environmental effects on the clusters were investigated by calculating the $J_{hor}$ and $J_{ver}$ action values as well as orbit parameters, such as eccentricity and inclination. We find that most of the RSG candidates in the clusters, except in RSGC1 and RSGC4, show circular disk-like motions and low inclination, as anticipated. 
In contrast, RSGC4 exhibits relatively low $J_{hor}$ and $J_{ver}$ values and high eccentricity.
RSGC1 has vertical motion with relatively high $J_{ver}$ values and high inclination compared with the other RSGCs.

Encounters with giant molecular clouds in the disk or the tidal force of bar potentials can induce increased vertical motion of clusters or disruption
of clusters. However, it remains unclear how the various differences in dynamical evolutionary properties among the clusters can result from 
the same environmental effects. 
For RSG candidates in RSGC4, it is more likely that the stars with radial motions toward the bar tip along the bar lane overshoot and then overlap with the stars traveling along the arms or gas streamers. We also considered binary or interacting clusters as the origin of the peculiar radial velocities and
action values of RSGC4. However, no apparent bimodalities were identified in the spatial positions and motions of the RSG candidates in RSGC4.

Overall, based on our analysis (the possibility of massive AGB stars, some binary systems of AGB-like stars, no runaway/walkaway stars, and dynamical
evolutionary properties), we tentatively conclude that RSGC4 may not be a genuine star cluster dominated by RSGs. 
Instead, it could be a complex assembly of stars, including RSGs and AGBs distributed along the line of sight at approximately the same distance.
These stars might originate from diverse environments, such as spiral arms and bar lanes.
The vertical motions of the RSG candidates in RSGC1 are also noteworthy.
Although RSGC4 might not be a star cluster, its dynamic properties, along with those of RSGC1 and other
RSGCs with disk-like motions, suggest a complicated and hierarchical secular evolution of star clusters in the Scutum complex region. 
In addition, our results caution against relying solely on density crowding to identify clusters in the disk and bulge regions.
Further investigation of the chemical abundances of RSGs and field massive stars, and detailed comparisons based on simulation, will provide keys to constraining whether RSGC4 is a true cluster and understanding the secular evolution in this region.

\begin{acknowledgments}
We are grateful to the anonymous referee for detailed comments and suggestions that greatly improved this paper.
S.H.C. acknowledges support from the National Research Foundation of Korea (NRF) grant funded by the Korea government (MSIT) (NRF-2021R1C1C2003511) and the Korea Astronomy and Space Science Institute under R\&D program (Project No. 2023-1-830-00) supervised by the Ministry of Science and ICT.
This work was supported by K-GMT Science Program (PID: GS-2018A-Q-418 and GS-2021A-Q-417) of Korea Astronomy and Space Science Institute (KASI).
This work used the Immersion Grating Infrared Spectrometer (IGRINS) that was developed under a collaboration between the University of Texas at Austin and the KASI with the financial support of the Mt. Cuba Astronomical Foundation, of the US National Science Foundation under grants AST-1229522 and AST-1702267, of the McDonald Observatory of the University of Texas at Austin, of the Korean GMT Project of KASI, and Gemini Observatory.
Based on observations obtained at the international Gemini Observatory, a program of NSF’s NOIRLab, which is managed by the Association of Universities for Research in Astronomy (AURA) under a cooperative agreement with the National Science Foundation on behalf of the Gemini Observatory partnership: the National Science Foundation (United States), National Research Council (Canada), Agencia Nacional de Investigaci\'{o}n y Desarrollo (Chile), Ministerio de Ciencia, Tecnolog\'{i}a e Innovaci\'{o}n (Argentina), Minist\'{e}rio da Ci\^{e}ncia, Tecnologia, Inova\c{c}\~{o}es e Comunica\c{c}\~{o}es (Brazil), and Korea Astronomy and Space Science Institute (Republic of Korea).
\end{acknowledgments}

\bibliographystyle{aasjournal}
\bibliography{reference}
\end{document}